\documentclass{aa}

\usepackage{graphicx}
\usepackage{txfonts}
\usepackage{lipsum}
\usepackage{subcaption}
\usepackage{lscape}
\usepackage{placeins}

\usepackage{xcolor}

\begin{document}

   \title{
   Electron transport in a radiation-dominated plasma\\
   Application to solar corona brightenings
   }

   \author{R. Duclous\inst{1} 
        \and V. Tikhonchuk\inst{2}
        \and M. Leboulanger\inst{1}
        \and X. Blanc\inst{3}
        \and A. S. Brun \inst{4}
        \and A. Strugarek \inst{4}
        }

   \institute{CEA, DAM, DIF, F-91297, Arpajon, France.
            \and
            Centre Lasers Intenses et Applications, Universit\'{e} de Bordeaux-CNRS-CEA, 33405 Talence, France.\\ 
             Extreme Light Infrastructure ERIC, ELI Beamlines Facility, 25241 Doln\'{i} B\v{r}e\v{z}any, Czech Republic.
             \and 
             Université Paris Cité and Sorbonne Université, CNRS, LJLL, F-75005 Paris, France.
             \and Université Paris-Saclay, Université Paris Cité, CEA, CNRS, AIM, 91191, Gif-sur-Yvette, France.\\
             \email{roland.duclous@cea.fr}
             }

   \date{Received 13 February 2026 / Accepted 31 July 2026}

  \abstract
   {Bremsstrahlung scattering of fast electrons on ions can be enhanced by the microwave radiation present
   in the solar corona. It can account for the electron diffusive transport along magnetic loops and
   high precipitation rates.
   This process can also dominate
   the transport of thermal electrons 
   confined in such loops,
   which calls for a dedicated study.
   }
   {We aim to study the influence of stimulated Bremsstrahlung scattering on the electron transport, 
   with a focus on the return current induced by the fast electron population trapped in magnetic loops. 
   From a more general standpoint, 
   transport coefficients need to be reevaluated in the radiation-dominated plasma, which is characterized by the stimulated action of 
   radiation on the Bremsstrahlung electron--ion collision frequency, down to thermal velocities.}
   {We developed a theoretical framework 
   for electron transport driven by a large bandwidth, bright, low-frequency part of the photon spectrum
   and computed a set of radiation-enhanced
   transport coefficients. Ultraviolet (UV), X-ray and extreme ultraviolet (XEUV)
   and hard X-ray signals from
   flares were reinterpreted in order to put maximum constraints on the 
   fast-electron model and its return-current component.
   These observations evidence anomalous resistivity, thermal-conduction inhibition, and high precipitation rates of fast electrons, which are 
   quantified by the model.}
   {
   Under the assumption of a near-isotropic solar microwave spectrum,
   stimulated Bremsstrahlung scattering provides the 
   enhancement of electron collision frequency
   needed to account for the 
   observed anomalous resistivity in flares.
   The anomalous resistivity systematically dominates over the classical resistivity
   in the flaring corona plasma by at least an order of magnitude. The runaway effect due to Coulomb collisions is suppressed.
   Thermal conduction is inhibited compared to the 
   Spitzer conduction, in agreement with coronal seismology of slow-mode waves.
   }
   {Stimulated Bremsstrahlung scattering is found to be a key collisional process
   in flaring events of the solar corona.
   It can explain
   the above loop-top hard X-ray signal due to the fast electrons
   and the measured electrical conductivity due to the thermal electrons.
   As a perspective, the corresponding transport coefficients 
   can be used in radiation MHD codes.
   To that aim, a simple model is proposed to self-consistently describe
   the transport coefficients and the infrared part of radiation spectrum.
   The radiation model could also be applied
   to stimulate large-angle electron scattering
   in the kinetic or hybrid models used 
   to study the reconnecting regions of the solar corona.
   }

               
   \maketitle
   \nolinenumbers
   
\section{Introduction}
\label{sec:intro}

High-resolution observations of brightening events in the solar atmosphere, categorized as flares and jets,
have improved statistics and our understanding of them. They have pointed out
the central role of magnetic reconnection in their rapid emergence \citep{Bahauddin2021NatAs}.
Enhanced spatial and temporal resolution have made it possible to observe, for instance,
the untangling of small-scale magnetic braids by reconnection in the solar corona \citep{Chitta2022}.

Aside from observations, magnetohydrodynamics (MHD) simulation stands as a 
prominent tool in the study of
the structure of the solar corona \citep{Klimchuk2015}, and in particular the role of flares and jets in its heating \citep{Raouafi2016}.
For instance, 3D MHD simulations helped us to recognize that magnetic flux emerging from the active Sun surface seeds prolific impulsive events, which can have a large-scale influence
\citep{Lu2024}. 
Dedicated MHD simulations have also been constrained by detailed observations. 
As an illustration, simulations of flare-reconnection sites were instrumented with synthetic spectroscopic diagnostics 
\citep{BROWNING2024}
and then compared to the typically observed quasiperiodic pulsations \citep{Kou2022}.

Appropriate transport coefficients relevant to MHD modeling are still unknown to a large extent, 
and classical ones, based on Coulomb collisions, do not suffice to explain many observations.
Among the possible processes, which could produce anomalous transport, 
turbulence is often considered to reconcile models with observations of the corona 
\citep{Minoshima2011ApJ73111M,Xu}.
Corresponding transport coefficients can be obtained \citep{Bian2016}.
Hence, several models of magnetic reconnection  \citep{Somov2013PartII},
or of electron transport in flare loops \citep{Xu},
retain as a constitutive assumption a negligible Spitzer resistivity compared to a presumed anomalous, collisionless resistivity.
Despite a considerable experimental effort devoted to turbulence characterization in flaring plasma \citep{Xie2024},
large uncertainties remain with respect to instability onset, while turbulence level 
needs to be calibrated for specific purposes.

Collisions between charged and neutral particles  bring another set of transport coefficients, although 
restricted  to the lowest temperature regions of the photosphere and chromosphere \citep{MacBride2022}.
 The corresponding transport coefficients \citep{Cowling1957}
are also associated with uncertainties related to abundances and out-of-equilibrium ionization models 
\citep{Nobrega2020}.

It was demonstrated that the radiation brightness temperature in the solar corona is strongly enhanced in the microwave spectral range
\citep{Landi2008}, in particular, in the flaring Sun conditions \citep{Duclous}. 
That leads to a modification in the effective electron--ion collision frequency compared to the Coulomb electron--ion collisions. This effect of stimulated Bremsstrahlung greatly enhances the diffusion of energetic electrons
(with energy about several tens of keV)
in the flaring loop and increases the precipitation rate.
In this study, we focused on the stimulated Bremsstrahlung scattering\footnote{The terminology
of stimulated Bremsstrahlung scattering refers to the 
electron scattering on an ion that combines with both the electron braking
due to Bremsstrahlung photon emission,
and the electron acceleration due to inverse Bremsstrahlung photon absorption.
This electron-scattering contribution, although usually neglected, is described by the Bremsstrahlung cross-section \citep{KochMotz1959}.
} 
of thermal electrons, which operates efficiently in 
flaring events in the solar corona. 
Stimulated Bremsstrahlung scattering of thermal electrons is described in Sect. \ref{sec:stimulatedscatt}.
It is demonstrated in Sect. \ref{sec:anomalousetrans} that this scattering process is dominant in the 
solar corona, leading to an anomalous transport.
Then, a model for the stimulated scattering frequency, relevant for the solar corona,
is proposed and discussed in Sect.
\ref{sec:modelScattFreq}. Its sensitivity with respect to the corona density and temperature gradients is studied in Sect.
\ref{sec:sensitivity}.
The stimulated transport coefficients 
are then computed in Sect. \ref{sec:transportcoeffs}.
They are used, in Sect. \ref{sec:application},
to reinterpret observations of solar flares. In Sect. \ref{sec:conclusion}, we conclude.

\section{Stimulated electron scattering by the low-frequency radiation}
\label{sec:stimulatedscatt}

In this section we introduce the stimulated Bremsstrahlung scattering frequency
used to compute transport coefficients, as well as the companion radiation model.
The 
equation for the radiation spectral intensity, $I_\nu$, 
can be written under the form \citep{bekefi,Rybicki,simoesKerr2017}
\begin{eqnarray}
\label{eq:mwrad}
\frac{d I_\nu}{d\tau_\nu } = B_\nu (T_e) - I_\nu \ , 
\end{eqnarray}
where $\displaystyle {B}_\nu (T_e)  $
is the Planckian spectral intensity at the electron temperature $T_e$,
$d\tau_\nu = \kappa_\nu ds$ is the optical depth for
inverse Bremsstrahlung 
along an elementary ray path $ds$, and $\kappa_\nu$ is the absorption coefficient at the photon frequency $\nu$.
For free--free transitions in hydrogen,
this absorption coefficient writes \citep{Rybicki}
\begin{equation}
\label{eq:opacityH}
\kappa_\nu = \frac{ \nu_{C,th}}{c}\frac{\nu_{pe}^2}{\nu^2} \frac{1}{N_r(\nu)}
\,,
\end{equation}
where  
$\nu_{pe}=(2\pi)^{-1} \sqrt{4\pi e^2 n_e /m_e}$
 is the plasma frequency ($e$ is the elementary charge, $m_e$ the electron mass,
 $n_{e}$ the electron density)
 and 
$N_r(\nu) = \sqrt{1-\nu_{pe}^2/\nu^2}$ 
 is the refraction index taken in the cold, unmagnetized limit approximation.
 Also,
$\nu_{C,th}=
\frac{4}{3} 
\left (2\pi \right )^{1/2}
n_e 
r_0^2c
\left( m_ec^2/k_B T_e \right)^{3/2} \ln \Lambda_\nu $ is the thermal electron--ion collision frequency.
Here, we denote the Boltzmann constant by $k_B$,
$c$ is the speed of light, and
 $r_0$ is the classical electron radius.
 Finally, the Coulomb logarithm is denoted by
 $\ln \Lambda_\nu= \ln \left( \varv_e^{th}/\nu b_{\rm min} \right)$, where  $\varv_e^{th}=(k_BT_e/m_e)^{1/2}$ is the electron thermal velocity, 
 $\displaystyle b_{\rm min} = \mbox{max} \left(e^2/k_B T_e, 
h / (2\pi m_e \varv_e^{th})
\right)$ is the minimum impact parameter for electron--ion collisions, and
$h$ is the Planck constant.\\

Equation \eqref{eq:mwrad} allows us to obtain the low-frequency component of the radiation field
needed to describe stimulated
Bremsstrahlung scattering
and can be solved in combination with the magnetohydrodynamics set of equations (see 
\citealt{Nordlund1982,Vogler2005}).
The useful part of this radiation field, which drives both the stimulated effect and absorption, is not intense
(that is, it does not contribute significantly to the radiation energy).
It corresponds to a large number of low-energy photons, spectrally located in an infrared divergence of the photon occupation number,
$ \displaystyle n_{\nu}  =  c^2 I_{\nu} / ( 2h\nu^3 ) $.
This feature was exploited in this work to describe the Bremsstrahlung 
cross-sections. Albeit there is a statistical offset between the stimulated emission and the absorption processes, both of them
cumulate the electron scattering contributions.
The probability of electron scattering on ions can then be greatly enhanced 
due to Bremsstrahlung.
The scattering frequencies due to stimulated emission, $\nu_{ei}^{B} $, and the one due to absorption,
$\nu_{ei}^{IB} $, are given in Appendix \ref{app:sbsfreq},
assuming a weakly anisotropic background radiation spectrum.
They can be approximated in the limit of low photon energy compared to the electron energy, 
$h\nu \ll \varepsilon_e$, yielding
a simple expression,
\begin{eqnarray}
\label{eq:simplenueis}
 \nu_{ei}^{SB} (\varv_e)  = 
 \nu_{ei}^{B} (\varv_e)  + 
 \nu_{ei}^{IB} (\varv_e)   \simeq  \frac{16}{3} \frac{\alpha_fZ ^2}{\beta_e} n_i r_0^2 c 
 \int_{\nu_{pe}} 
 d\nu \, (1+2n_\nu) / \nu 
 \ ,
\end{eqnarray}
 where  $\alpha_f$ is the fine structure constant, $n_i$ is the ion density, $Z$ is the atomic number, and
 $\beta_e=\varv_e/c$ is the normalized electron velocity.

\begin{figure}[h!]
    \centering
    \includegraphics[width=\hsize]{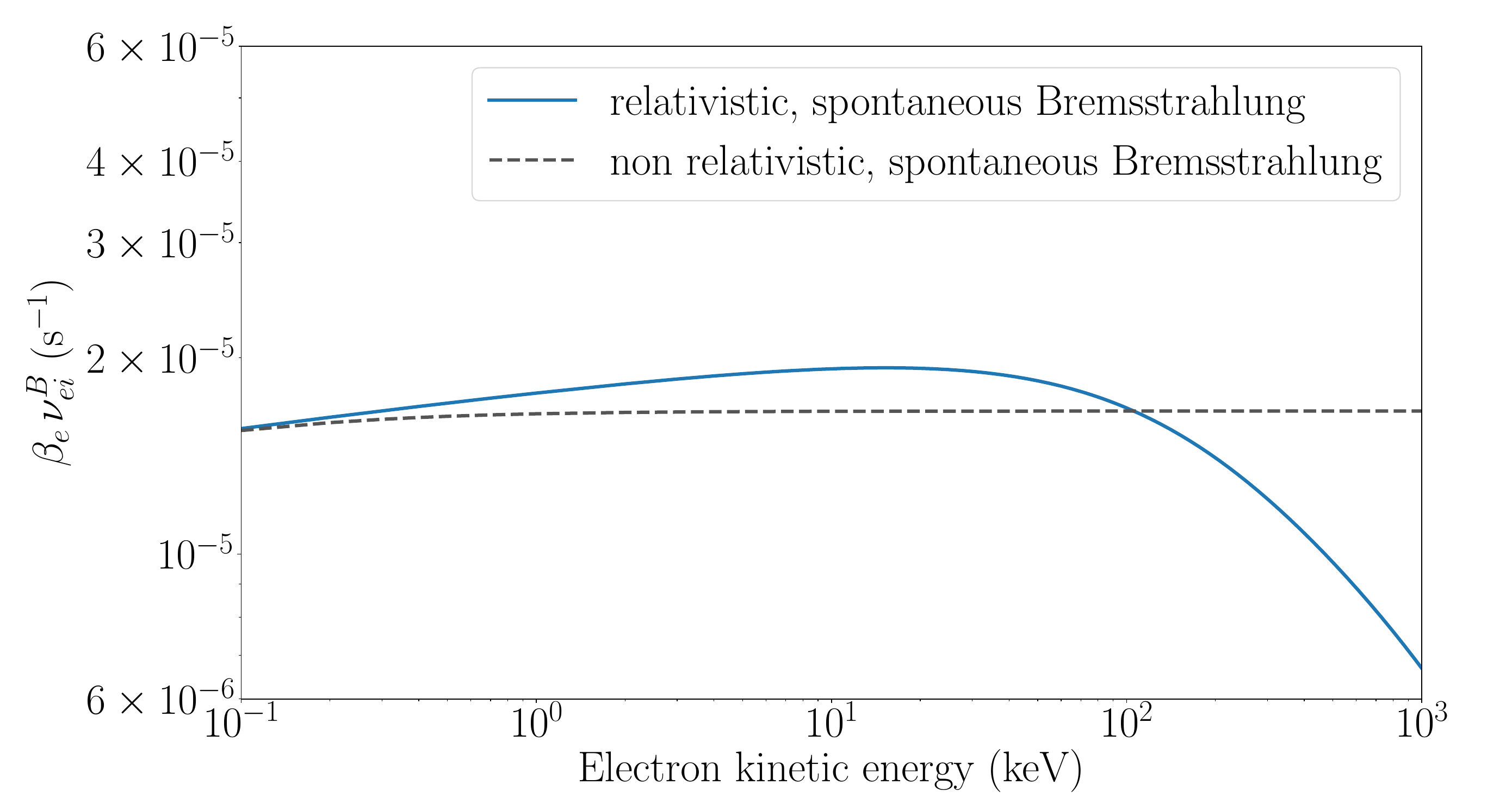}
       \caption{ Normalized collision frequency, $\ \beta_e  \nu^B_{ei} $, as a function of the electron kinetic energy, $\varepsilon_e$, 
       for a fully ionized hydrogen plasma with $n_e=10^{10} \, \mbox{cm}^{-3}$. 
       Only spontaneous Bremsstrahlung emission was considered, neglecting the radiation field. 
       The collision frequency, $\nu^B_{ei} $, was obtained by integrating the relativistic 
       doubly differential cross-section in photon energy and electron direction \citep{Racah1934,mccormick1956}. 
       The nonrelativistic approximation was obtained from  
       the Bethe-Heitler differential cross-section.}
          \label{fig:betanu}
\end{figure}

The Bremsstrahlung scattering frequency scales as the inverse of the electron velocity, 
 which is much less pronounced than the cubic scaling
 with electron velocity in the
 Coulomb collision frequencies.
The electron mean-free-path thus becomes independent of velocity.
This scaling is valid for electron energies below $100$ keV,
where the nonrelativistic approximation is accurate (with a maximum difference of $\sim 10 \%$).
This is illustrated  for the spontaneous Bremsstrahlung process in Fig. \ref{fig:betanu}, 
which shows $\ \beta_e  \nu^B_{ei}$, assuming $n_\nu \ll 1$, as a function of the electron kinetic energy.

We introduce the effective
scattering frequency due to stimulated Bremsstrahlung, 
as an average over the electron distribution function
 \begin{eqnarray}
 \label{eq:avrgnueiSB}
 \overline{\nu}_{ei}^{SB} \equiv \frac{3}{4\pi} \frac{n_ek_BT_e}{m_e} \left [  \int_{0}^{\infty} d \varv_e \frac{\varv_e^4 M_e}{\nu_{ei}^{SB}(\varv_e)} \right ]^{-1} \simeq \frac{3}{8} \sqrt{ \frac{\pi}{2} } \nu_{ei}^{SB}(\varv_e^{th}) 
 \ ,
 \end{eqnarray}
where
 \begin{eqnarray}
 M_e \equiv n_e 
 \left( \frac{m_e}{2\pi k_b T_e} \right)^{3/2} \; 
\text{exp}\left(- \frac{m_e \varv_e^2}{2k_b T_e}\right)
 \end{eqnarray}
 is the Maxwellian distribution.
This effective scattering frequency was used 
in the transport coefficients due to stimulated Bremsstrahlung, given in Sect. \ref{sec:transportcoeffs}.
 The magnitude of the scattering frequency was determined by 
the two infrared divergences in Eq. \eqref{eq:simplenueis}.
If $n_\nu \ll 1$ at $\nu \sim \nu_{pe}$, the spontaneous Bremsstrahlung operates,
and the frequency integral is logarithmically divergent. 
It can be estimated as $\ln(\nu_{cut}/\nu_{pe})$, with $\nu_{cut} \sim \varepsilon_e/h$. 
In this limit, 
the Bremsstrahlung frequency is smaller than the electron--ion Coulomb collision frequency \citep{Spitzer1953,Kuritsyn2006,huba2016nrl}:
\begin{equation}
\label{eq:nu_ei_coulomb_average}
\overline{\nu}^{C,SH} = \frac{4}{3} (2\pi)^{1/2} n_i r_0^2cZ^2 \left ( \frac{m_ec^2}{k_BT_e} \right )^{3/2} \ln \Lambda_{ei} \ .
\end{equation}
Here, we are interested in
the reverse ordering: $n_\nu \gg 1$ at $\nu \sim \nu_{pe}$.
This limit corresponds to the stimulated Bremsstrahlung, where
the infrared divergence in the photon occupation number also operates. 
It is shown in the next section that the stimulated Bremsstrahlung frequency can be much higher 
than the Coulomb collision frequency for electrons in the solar corona.

For the sake of completeness, it should be pointed out
that the absorption coefficient \eqref{eq:opacityH} depends on the
Coulomb scattering frequency but not on the stimulated Bremsstrahlung scattering frequency. 
This is because Eq. \eqref{eq:mwrad}  for radiation transfer
already incorporates the stimulated effect.

\section{Anomalous electron transport in the solar corona} 
\label{sec:anomalousetrans}
\begin{figure}[h!]
    \includegraphics[width=\hsize]{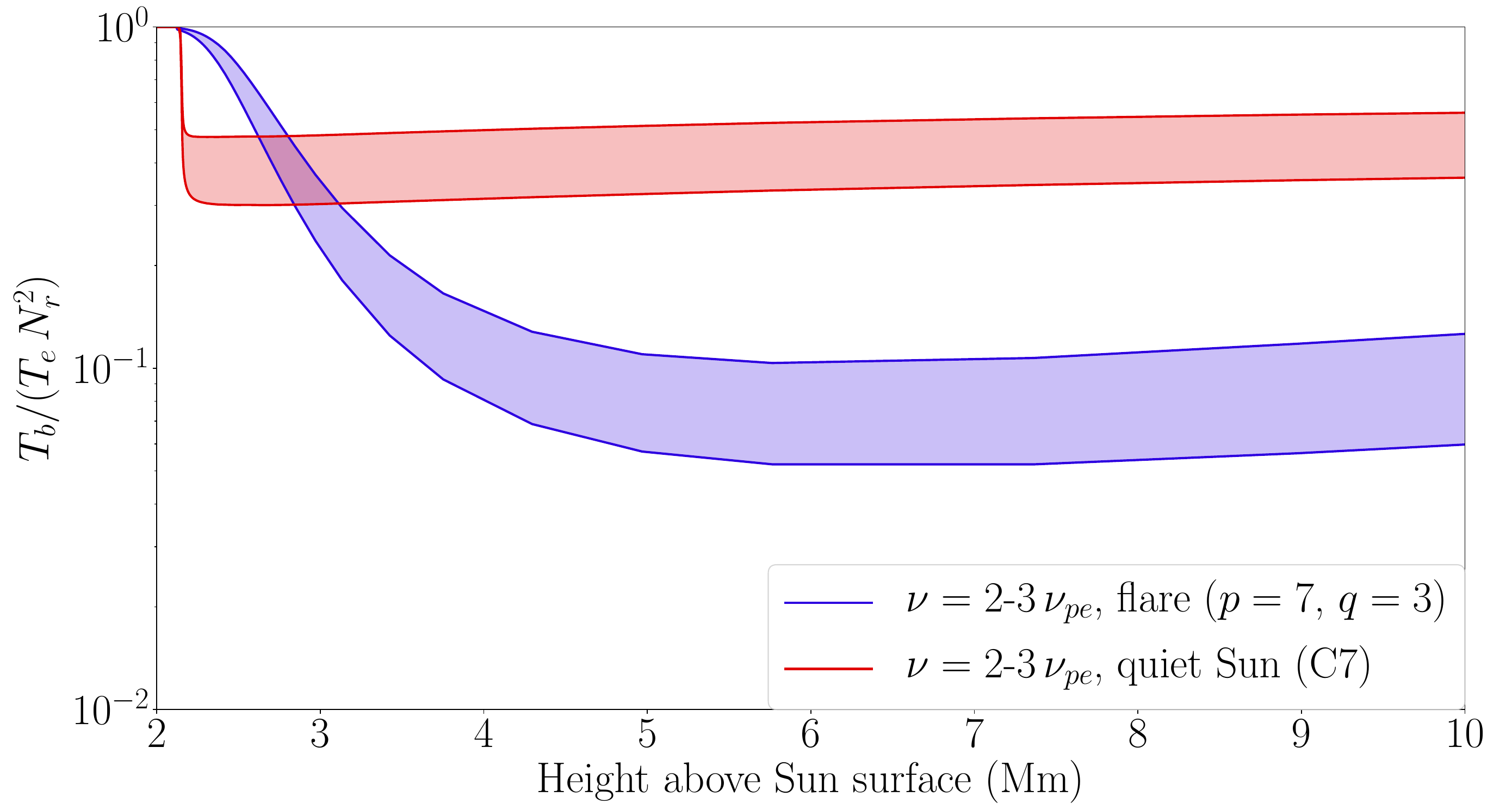}
       \caption{Normalized brightness temperature of photons
       with a frequency of $\nu=2$-$3 \, \nu_{pe}$
       for the quiet (red) and flare (blue) solar atmospheres. 
       Details of calculations are given in Appendix \ref{app:parametrization}.
       }
          \label{fig:hprofileTbTe}
\end{figure}

The solar corona provides plasma conditions favorable to a prominent stimulated Bremsstrahlung scattering of thermal electrons.
The infrared divergence of the radiation spectrum falls in the microwave spectral range and drives 
thermal electron transport due to its high brightness.
We computed synthetic microwave-radiation spectra in the corona,
solving  Eq. \eqref{eq:mwrad} numerically in the Sun atmosphere from its surface,  with the same method as \cite{Duclous}.
Two models of solar atmosphere were considered that are representative of either the quiet Sun or a flare.
For both model atmospheres, Fig. \ref{fig:hprofileTbTe} provides evidence of
a high brightness temperature, 
$\displaystyle T_b(\nu) \equiv c^2 I_\nu N_r^2(\nu) \left   / \, \left  ( 2\nu^2 k_B \right ) \right .  $, 
that approaches the corona electron temperature.
The ratio $T_b/T_e$ is about 40 \% in the quiet Sun and about 10 \% in a flare.
The shadowed zones reveal a large number of photons near the plasma cutoff,
$\nu=2$-$3 \, \nu_{pe}$, which are in near-thermal equilibrium with electrons in the corona.
 
\begin{figure}[h!]
    \includegraphics[width=\hsize]{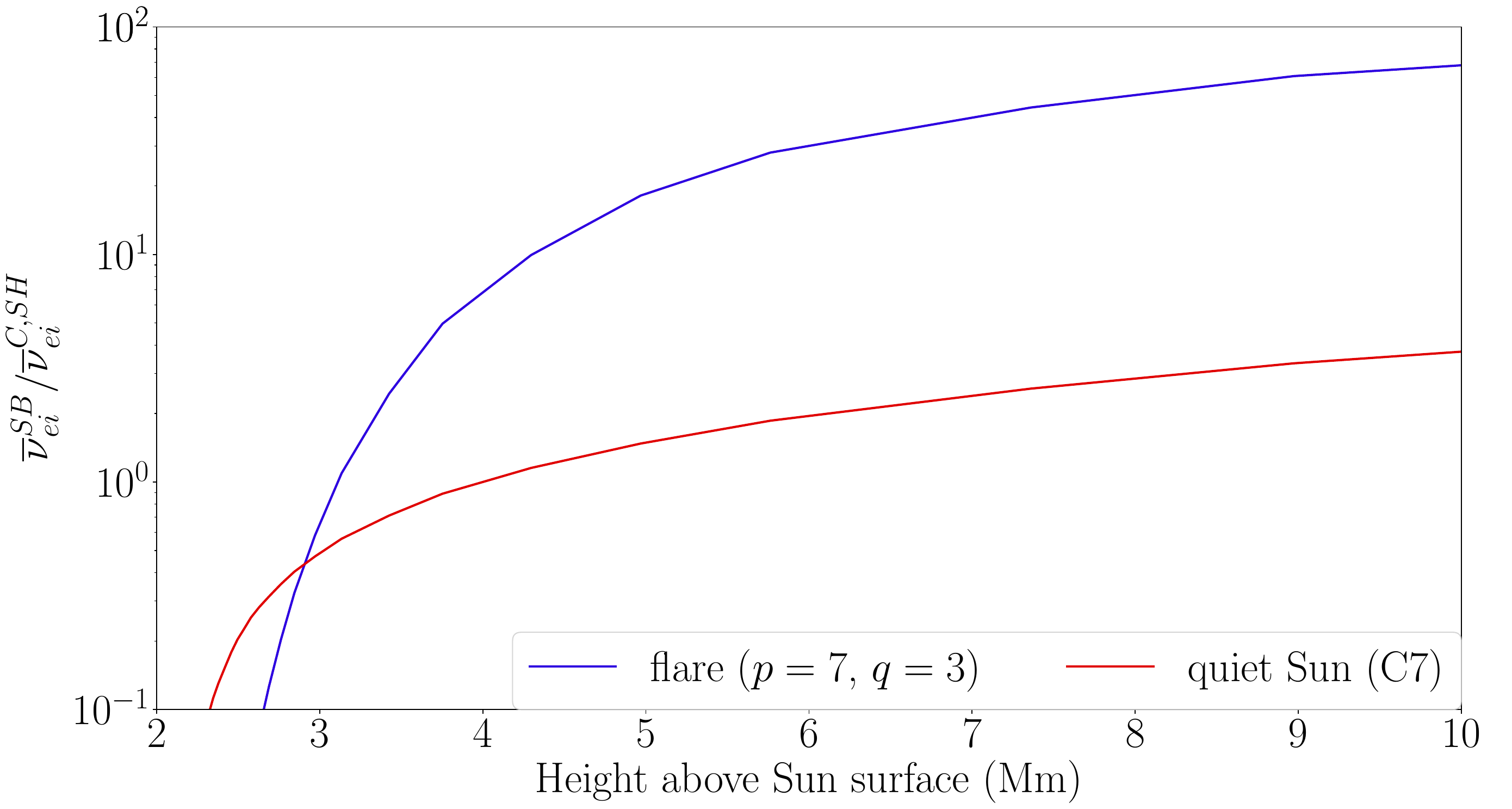}
       \caption{Effective scattering frequency due to stimulated Bremsstrahlung, $\overline{\nu}_{ei}^{SB}$, compared to the
        scattering frequency due to Coulomb collisions in fully ionized hydrogen,
in the solar transition layer and corona, for the quiet-Sun and flare atmospheres 
detailed in Appendix \ref{app:parametrization}.
       }
          \label{fig:hprofileCollFreqP7Q3}
\end{figure}

The synthetic corona's microwave spectra are used to
evaluate the average scattering frequency \eqref{eq:avrgnueiSB}.
 It is compared to the Coulomb scattering frequency
 in Fig. \ref{fig:hprofileCollFreqP7Q3}
 for the quiet and flaring solar corona. 
In the flaring plasma, the scattering frequency due to stimulated Bremsstrahlung is about
10 to 100 times higher than the Coulomb scattering frequency. 
In the quiet-Sun corona plasma, it is about one to four times higher.\\

Such a significant enhancement of scattering by stimulated Bremsstrahlung 
leads to
a strong reduction of the electrical and thermal conductivities.
The MHD equations related to electron transport are revised in Sect. \ref{sec:transportcoeffs}.
Presumably, the modification can bring a very different flare dynamics than the one classically predicted.
Observational manifestations are discussed in Sect. \ref{sec:application}.

\section{Stimulated Bremsstrahlung scattering frequency}
\label{sec:modelScattFreq}

\begin{figure*}[h!]
    \resizebox{\hsize}{!}
             {\includegraphics{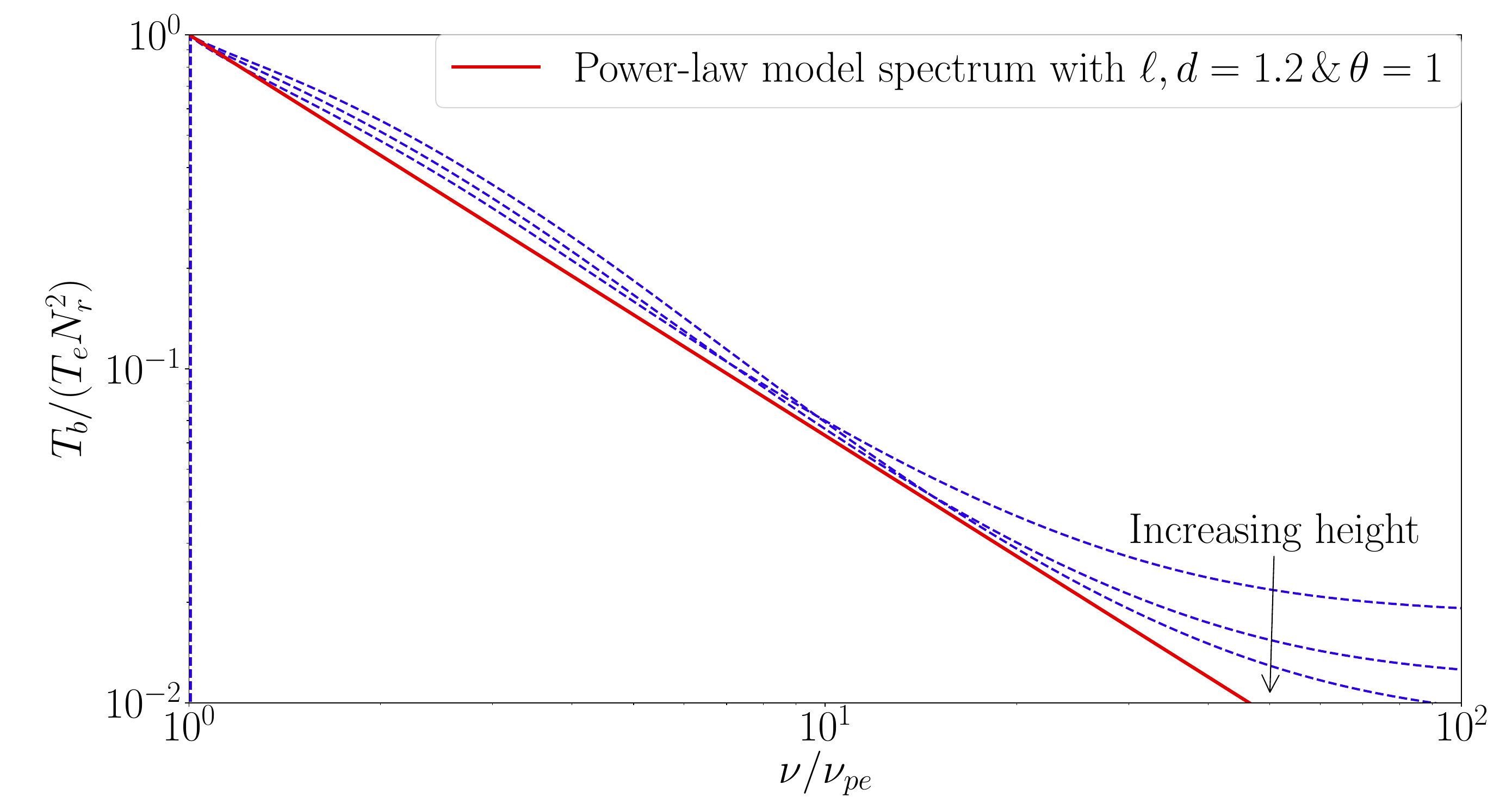}
              \includegraphics{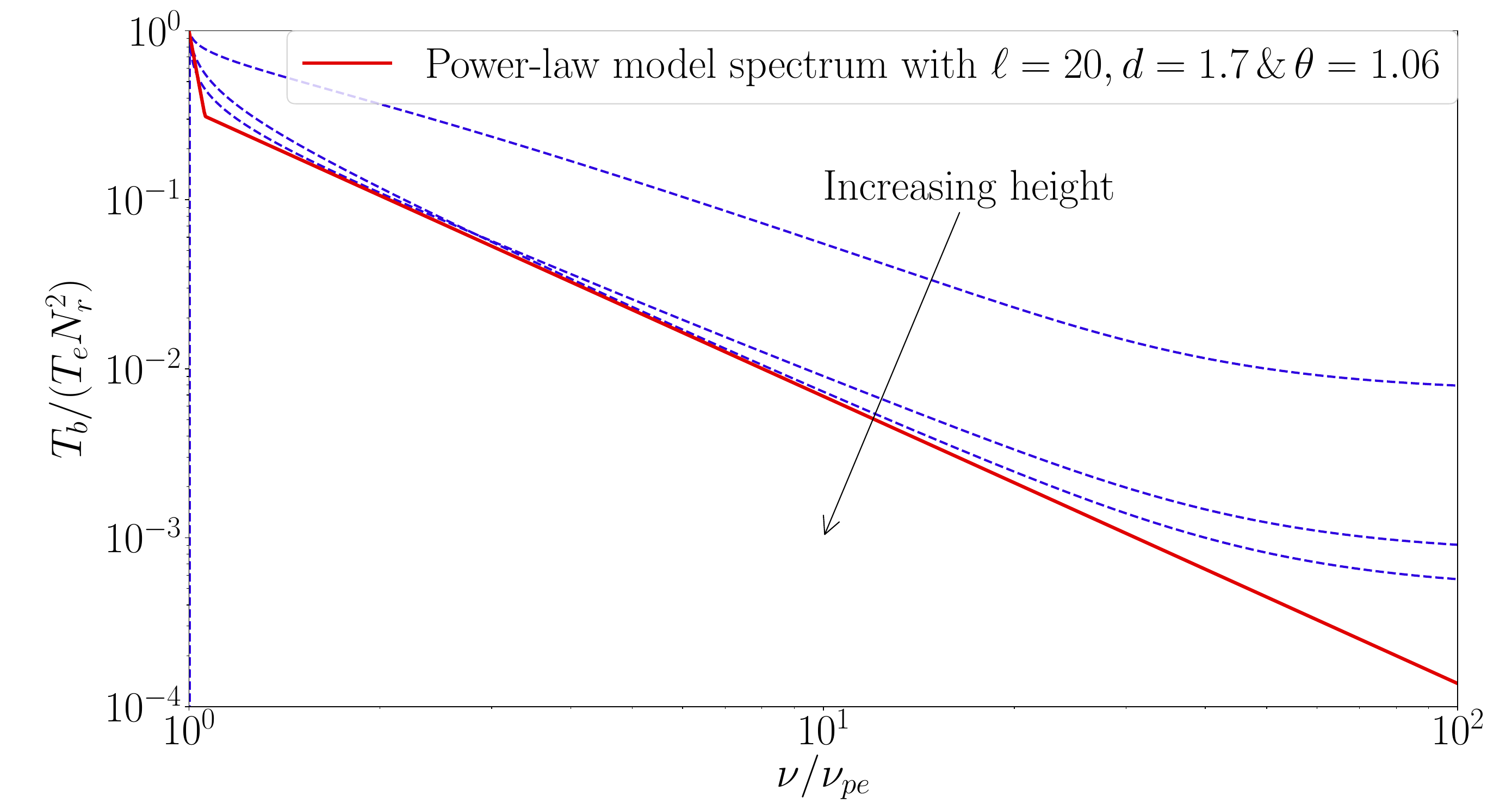}}
       \caption{Low-frequency corona radiation spectra at different heights above the Sun's surface, $h=3,5,9$ Mm
       (dashed curves).
       Different solar atmospheres were considered, namely, the C7 solar atmosphere from 
       \citet[\textit{left}]{Avrett_2008}
       and its modification 
       from Appendix \ref{app:parametrization} with $p=7$ and $q=3$ for a flare ({\it right}). 
       The model radiation spectrum of Eq. \eqref{eq:powerlaw} is also shown with solid red lines.
       The quiet-Sun radiation spectrum, on the left-hand side, was fitted  at $h=9$ Mm  with this model using $\ell=d=1.2$ and $\theta=1$, which gives $\ell^*=1.2$.
       The flaring-Sun radiation spectrum, on the right-hand side, was fitted at same height using $\ell=20$, $d=1.7$ and $\theta=1.06$, which gives $\ell^*=6$.
       }
          \label{fig:TeTB}
 \end{figure*}

Stimulated Bremsstrahlung relies on a background  microwave radiation in the solar corona.
Assuming that the plasma is optically thin, the radiation field can be approximated with a Rayleigh-Jeans distribution
 with the temperature, $T_{rad}$, being that of the Sun's surface. The
 occupation number 
 in the spectral range $\nu_{pe} < \nu < k_B T_{rad}/h$  can thus be approximated as 
 $n_{\nu} \sim k_BT_{rad}/(h\nu)$ in Eq. \eqref{eq:simplenueis}.
  The stimulated Bremsstrahlung collision frequency can be estimated as
\begin{eqnarray}
\left . \nu_{ei}^{SB} \right |_{thin} \left( \varv_e \right) =
\frac{32}{3} \frac{\alpha_fZ ^2}{\beta_e} n_i r_0^2 c \frac{k_BT_{rad}}{h\nu_{pe}} \ . 
\end{eqnarray} 
It can be compared to
 the electron--ion Coulomb collision frequency  \citep{Jackson1962},
 \begin{equation}
\label{eq:nu_ei_coulomb}
\nu_{ei}^C \left( \varv_e \right)  =  4 \pi n_i r_0^2 c Z^2
\beta_e^{-3} 
\ln \Lambda_{ei} \ ,
\end{equation}
 which gives the frequency ratio
 \begin{equation}
 \label{eq:BremDominates}
 \frac{
\left . \nu_{ei}^{SB}  \right |_{ thin } 
 (\varv_e)}{\nu^C_{ei}(\varv_e)} \mathrel{ \underset{{n}_\nu \gg 1}{\simeq}} 
 \frac{8}{3 \pi}
 \frac{ \alpha_f \beta_e^2 }{\ln \Lambda_{ei}}
 \frac{k_B T_{rad}}{h\nu_{pe}} \ .
 \end{equation}
 Although this expression is useful for qualitative estimates, it does not apply to solar corona,
 where the plasma is translucent, with significant  optical thickness
to the low-frequency photons  \citep{Landi2008}.
 The microwave spectrum follows a power-law with high brightness temperature, typically much higher than $T_{rad}$.
The power-law spectrum eventually connects, at higher frequencies, to the Rayleigh-Jeans spectrum.
 In the other limit, $\nu \sim \nu_{pe}$,
Eq. \eqref{eq:mwrad} imposes 
$I_\nu \sim  B_\nu (T_e)$ or, 
equivalently, $T_b(\nu) \, \left / \, N_r^{2}(\nu) \right . \sim  T_e$.
 Then, the low-frequency radiation spectrum in the solar corona can be parametrized as
 \begin{eqnarray}
 \label{eq:powerlaw}
 T_b(\nu) \, \left /  \, N_r^{2}(\nu) \right . =
\left \{ 
\begin{array}[!h]{cc}
\displaystyle T_{e} \left ( \nu_{pe} / \nu \right)^{\ell} \, \ , &   \ \nu  < \theta \nu_{pe}  \\ 
\displaystyle  \frac{T_{e} }{ \theta^l} \left ( \theta \nu_{pe} / \nu \right)^d \, \ ,  &  \ \nu > \theta \nu_{pe}  
  \end{array}
\right .  
, \end{eqnarray}
where $\ell$ and $d$ are positive parameters and $\theta \geq 1$. 
Here, the plasma refraction index, $N_r$, accounts for the evanescent radiation field near plasma frequency
\citep{bekefi}.
The radiation spectrum \eqref{eq:powerlaw} agrees with
the measured microwave spectra \citep{Landi2008} and
with  the synthetic corona's microwave-radiation spectra
obtained from Eq. \eqref{eq:mwrad}.
Fitted synthetic spectra are shown in Fig. \ref{fig:TeTB} at various corona heights, for both 
the C7 quiet-Sun atmosphere from \cite{Avrett_2008}, and the 
parametrized flare corona of Appendix \ref{app:parametrization}. 
In these spectra, the brightness temperature in the low-photon-frequency range, near plasma frequency,
is clearly related to the corona electron temperature.
Figure \ref{fig:TeTB} also suggests a self-similar behavior of the microwave spectra in the solar corona.
It is noteworthy that the low-frequency radiation model \eqref{eq:powerlaw} can be inserted in the stimulated Bremsstrahlung scattering frequency \eqref{eq:simplenueis}, yielding
 \begin{equation}
\label{eq:BremPowerLaw}
  \nu^{SB}_{ei}  (\varv_e) \mathrel{ \underset{{n}_\nu \gg 1}{\simeq}}  \frac{32}{3} \frac{\alpha_f Z^2}{\beta_e} n_i r_0^2 c \frac{k_BT_{e}}{h\nu_{pe}} \frac{1}{1+\ell^*} \ ,
\end{equation}
where $\ell^*$ is an average spectral slope, which
accounts for a difference between the electron and radiation brightness temperatures and
satisfies the convex combination
\begin{eqnarray}
 1/(1+\ell^*)  = \left . \left [ 1-1/\theta^{1+\ell} \right ] \, \right /(1+\ell) + \left . \left [ 1/\theta^{1+\ell} \right ] \, \right / (1+d) \ .
 \end{eqnarray}
The frequency ratio becomes
 \begin{equation}
 \label{eq:BremDominates2}
 \frac{\nu^{SB}_{ei}  (\varv_e)}{\nu^C_{ei}(\varv_e)} \mathrel{ \underset{{n}_\nu \gg 1}{\simeq}} 
 \frac{8}{3 \pi}
 \frac{ \alpha_f \beta_e^2 }{\ln \Lambda_{ei}}
 \frac{k_B T_{e}}{h\nu_{pe}}  \frac{1}{1+\ell^*}  \ .
 \end{equation}
 Eq. \eqref{eq:BremPowerLaw} relies on an enhanced microwave spectrum 
 (characterized by the brightness temperature for photons with $\nu \sim \nu_{pe}$ approaching the electron temperature,
 which is much higher than the bulk radiation temperature, $T_{rad}$).
 This implies that the plasma optical depth,
 $\tau_\nu = \int ds \, \kappa_\nu$,
 should be sufficiently large for the low-frequency photons.
 The enhancement effect on the collision frequency, compared to the optically thin condition,
 can be evaluated as $\tau_{\nu} T_e/T_{rad}$.
 Hence, a figure of merit, denoted by $F_0$ below, can be put forward to 
 evaluate the efficiency of stimulated Bremsstrahlung scattering of thermal electrons.
 For $\tau_\nu <1$ and photon frequencies near the plasma frequency, $\nu \sim 3 \nu_{pe}$, it can be defined as a product:
 \begin{eqnarray}
 \label{eq:F0}
 F_0 \equiv  \tau_{\nu}  (T_e/T_{rad}) \, \times \, \left .  \nu^{SB}_{ei}  \right |_{thin} (\varv_e^{th}) / \nu^C_{ei}(\varv_e^{th})  \propto n_i^{1/2}T_e^{1/2} L \ ,
 \end{eqnarray}
 where $L$ is the plasma size.
 The  scaling of Eq. \eqref{eq:F0} is the net result of competing dependencies in
 the frequency ratio $ 
 (T_e/T_{rad}) \times 
  \left .  \nu^{SB}_{ei}  \right |_{thin}
 (\varv_e^{th}) / \nu^C_{ei}(\varv_e^{th}) \propto n_i^{-1/2} T_e^2$
 and the optical depth
 $\tau_\nu \propto n_iT_e^{-3/2} L$.
 The figure of merit, $F_0$, can be estimated using $\tau_\nu \sim L \, \nu_{C,th} / (9 c)$, 
 for an idealized
corona, represented by a homogeneous plasma slab of 
$L=10$ Mm in height above the Sun's surface. 
In a quiet-Sun corona with $n_i=3.6 \times 10^8$ cm$^{-3}$,  $T_e=1$ MK, we obtain $F_0 \sim 0.6$; stimulated Bremsstrahlung 
scattering and Coulomb collisions are of equal importance.
By contrast, for a flare  with $n_i= 10^{10}$ cm$^{-3}$, $T_e=15$ MK,
a much higher figure of merit is obtained, $F_0 \sim 12 \gg 1 $.

\section{Model sensitivity to density and temperature gradients in the solar corona}
\label{sec:sensitivity}

The flaring corona is highly dynamic, plasma conditions 
can also vary a lot from one flare to another, which modifies the microwave radiation spectra.
In order to take that variability into account,
in the computation of the stimulated Bremsstrahlung scattering frequency, 
we considered various electron temperature and density gradients; namely small, intermediate, and large gradients,
assuming a fixed temperature and density at the loop top.
 The flare corona parametrization noted in Appendix \ref{app:parametrization} was considered, 
 setting the gradients of electron density and temperature with the free parameters $p,q \in \{ 3,7,20\}$.
 
\begin{figure}[h!]
    \includegraphics[width=\hsize]{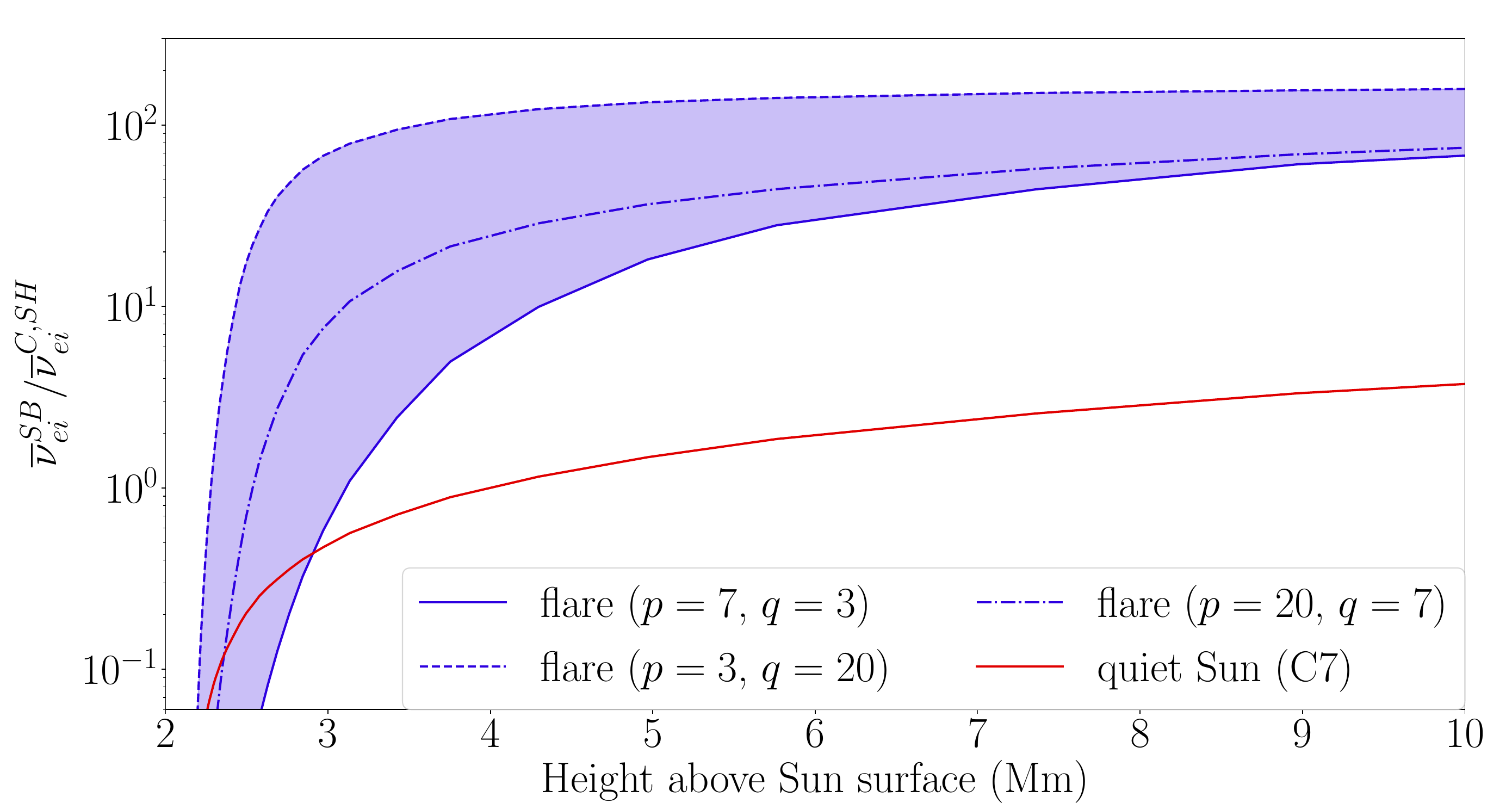}
       \caption{Ratio of
       average scattering frequency due to stimulated Bremsstrahlung, $\overline{\nu}_{ei}^{SB}$, and 
       average scattering frequency due to Coulomb collisions,
in the solar transition layer and corona.
Details of calculations are given in  Appendix \ref{app:parametrization}.
       }
          \label{fig:hprofileCollFreq}
\end{figure}

\begin{figure}[h!]
    \includegraphics[width=\hsize]{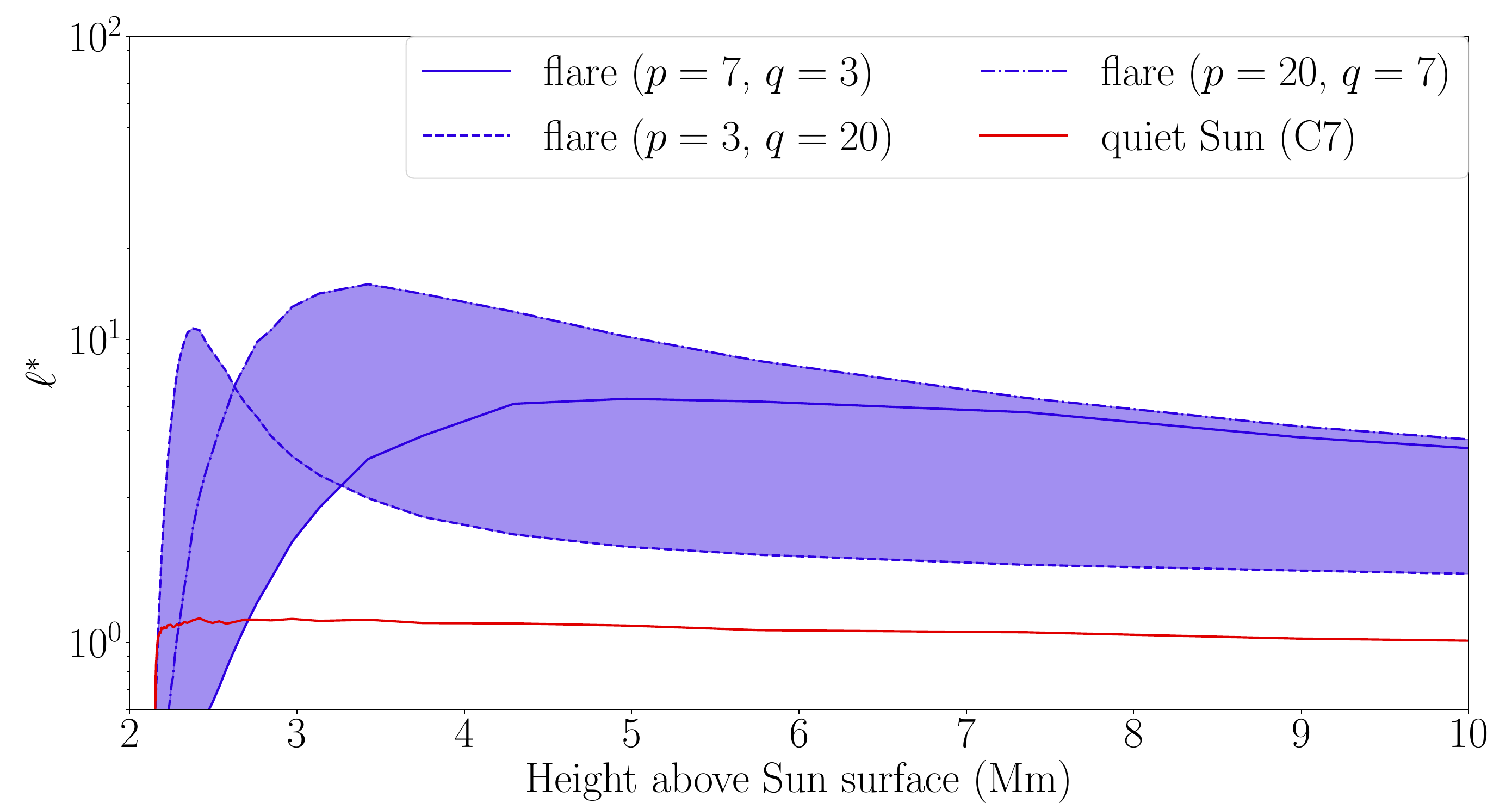}
       \caption{Average spectral index, $\ell^*$, as a function of height, in the solar transition layer and corona.
It is shown for the C7 quiet-Sun model from \cite{Avrett_2008} in red.
It lies in the blue shaded area for the parametrized flare model of Appendix \ref{app:parametrization} with $p,q \in \{ 3,7,20 \}$.
       }
          \label{fig:hprofileBstar}
\end{figure}

The variability in the flare atmosphere was assessed by considering
the average  scattering frequency due to stimulated Bremsstrahlung, $\overline{\nu}_{ei}^{SB}$,
which is described in Sect. \ref{sec:anomalousetrans} as being obtained
from synthetic radiation spectra and was compared to a Coulomb scattering frequency.
The dependence of the scattering-frequency ratio on the density and temperature gradients 
is shown in Fig. \ref{fig:hprofileCollFreq} for solar-flare conditions.
The influence on gradients proves to be small, as the blue shaded area in this figure is narrow.
This area contains the possible scattering ratio obtained by varying the density and temperature gradients,
resulting in various coronal flare atmospheres.
Similarly, the model scattering frequency in Eq. \eqref{eq:BremPowerLaw} 
has low sensitivity to the density and temperature gradients.
This is shown in Fig. \ref{fig:hprofileBstar}, which demonstrates a weak dependance of
the average spectral slope, $\ell^*$.
There, $\ell^*$ is obtained from Eq. \eqref{eq:BremPowerLaw}, inserting the
 synthetic corona's radiation spectra in  $\nu_{ei}^{SB}(\varv_{e}^{th})$.
We find that a conservative choice, $\ell^* \sim 6$, can be considered for a flare,
 while for the quiet Sun, $\ell^* \sim 1$ is appropriate. 
 Note that these values can also be obtained 
 by directly fitting the radiation spectra; see Fig. \ref{fig:TeTB}.

Hence, stimulated Bremsstrahlung dominates in solar flares. 
Its effect on the thermal electron transport is characterized in the next section.

\section{Transport coefficients for a radiation MHD model}
\label{sec:transportcoeffs}

Let us consider, for the sake of simplicity, a radiation-dominated plasma that satisfies the following frequency ordering:
$
\nu^{SB}_{ei}(\varv_e) \gg \nu^{C}_{ei}(\varv_e), \nu^{C}_{ee}(\varv_e)$ for $\varv_e \gtrsim \varv_e^{th}$.
In such a plasma, the transport coefficients are driven by electron scattering on ions due to stimulated Bremsstrahlung. 
Their computation follows the approach of \cite{Spitzer1953}. It is based 
on the weak-anisotropy assumptions and thus can be 
performed within the P1 approximation 
of the Vlasov equation for electrons in a fixed frame. 
In this approximation, only the zeroth-order and first-order spherical harmonics of the electron
distribution function, $ \mbox{f}_e$, were considered:
   \begin{displaymath}
 \mbox{f}_{e,0}  \equiv  \frac{1}{4\pi} \int_{S^2} d\Omega_e  \mbox{f}_e  \, , \;
{\bf f_{e,1}}  \equiv  \frac{3}{4\pi} \int_{S^2} d\Omega_e {\bf \Omega_e}  \mbox{f}_e \ .
   \end{displaymath}
They satisfy the following condition: $\left | {\bf  f_{e,1}} \right | / \mbox{f}_{e,0} \ll 1 $. 
At first order in the Chapman-Enskog expansion, the solution is the Maxwellian distribution,
$ \mbox{f}_{e,0} = M_e$.
At the next order, the first-order spherical harmonic satisfies the stationary equation
\begin{eqnarray}
\label{eq:baseResistivity}
\varv_e \nabla_{\bf x} M_e -  \frac{e{\bf E}}{m_e} \frac{\partial M_e}{\partial \varv_e}
- \frac{e {\bf B}}{m_ec}
 \, \times \, 
{\bf f_{e,1} } = - \nu^{SB}_{ei}(\varv_e) \, {\bf f_{e,1} } \ ,
\end{eqnarray}
where ${\bf E}$ and ${\bf B}$ are the electric and magnetic fields, here assumed quasi-static.

\begin{figure*}[h!]
    \resizebox{\hsize}{!}
             {\includegraphics{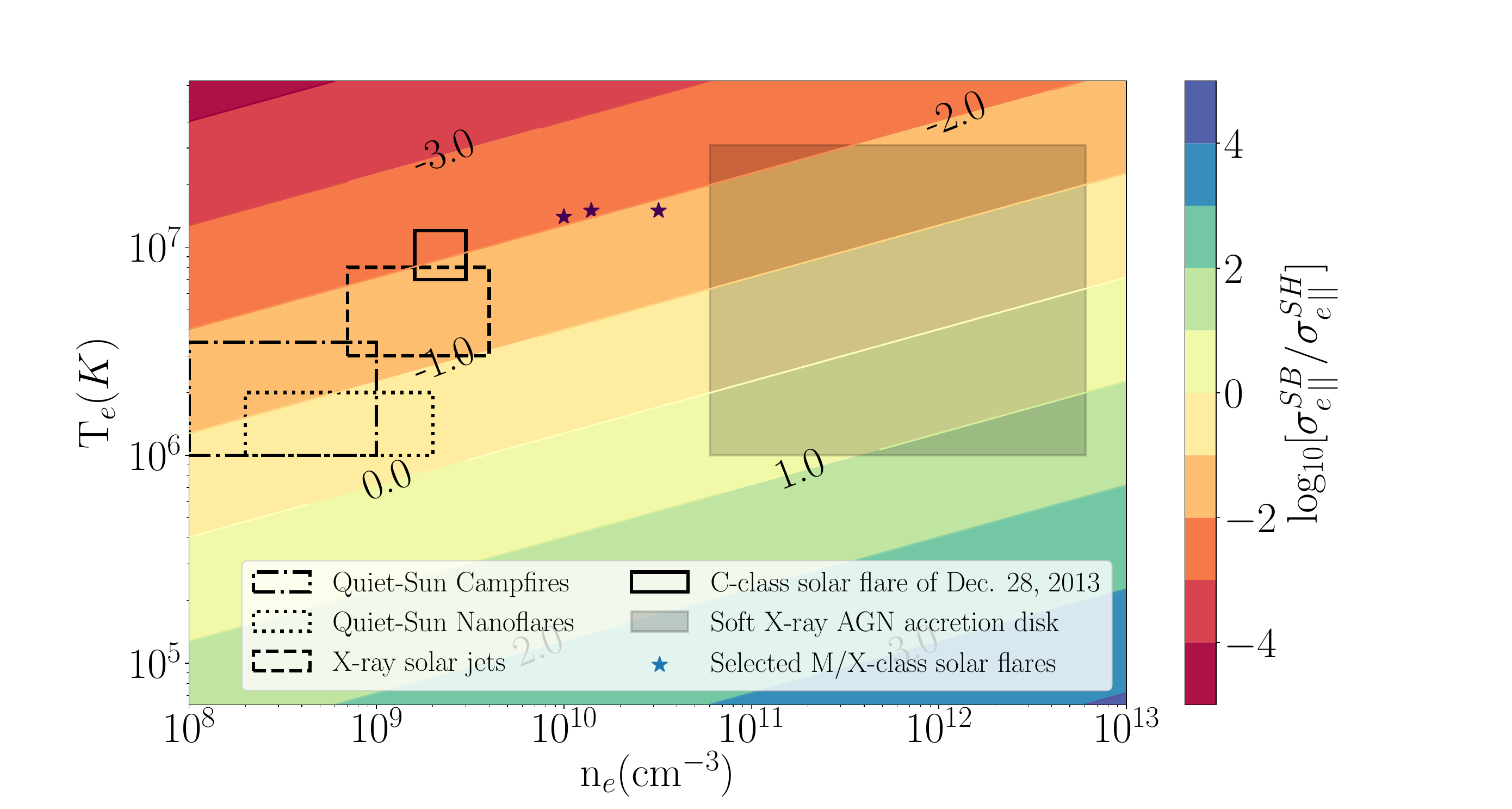}
              \includegraphics{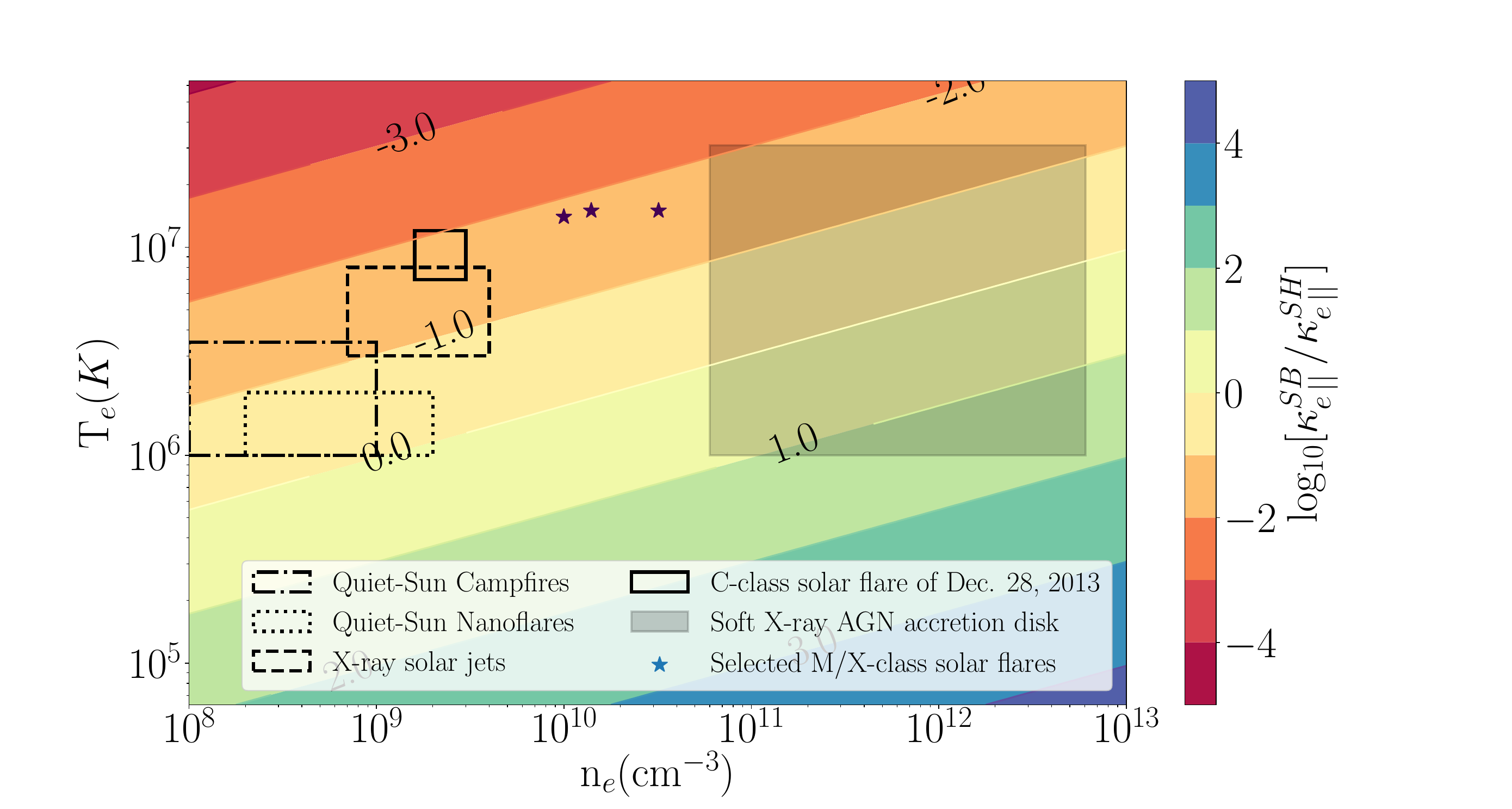}}
       \caption{Electrical ({\it left}) and thermal ({\it right}) conductivities due to stimulated Bremsstrahlung, normalized to the corresponding Spitzer conductivities, shown in logarithmic scale as a function of the plasma 
       density and temperature. 
       To evaluate the conductivities, the frequency ratio 
       $  \overline{\nu}_{ei}^{SB}  / \overline{\nu}^{C,SH} $ is assumed to scale as $T_e^2 \, n_e^{-1/2}$
       (see Eq. \eqref{eq:BremDominates2}), while its value at $T_e=15$ MK and $n_e=10^{10}$ $\mbox{cm}^{-3}$ is conservatively set to 70 (see Fig. \ref{fig:hprofileCollFreq}).
       The average spectral slope is considered constant and set at the conservative value $\ell^* \sim 6$
       (see Fig.  \ref{fig:hprofileBstar}).
       The three flares listed in Table \ref{table:flareparameters} are indicated with stars, while
       the flares analyzed in Sect. \ref{subsec:thermalcond} are indicated with
       bold rectangles.
        Typical values relevant for X-ray jets \citep{shen2021}, quiet-Sun nanoflares and campfires \citep{Podladchikova2025arXiv}
        are also indicated with rectangles.
        Soft X-ray-emitting accretion disks of active galactic nuclei \citep{Igarashi2024} are shown
        with gray shaded rectangles.
       }
          \label{fig:conductivity_ratio}
 \end{figure*}

An explicit expression for ${\bf f_{e,1}}$ can be deduced from Eq. \eqref{eq:baseResistivity}
to describe the magnetized transport with a similar approach to that of \cite{Braginskii1965};
see Appendix \ref{app:magelectras}.
In the particular cases of zero magnetic field or transport along the magnetic-field lines,
which are considered in Sect. \ref{sec:application}, the integration of Eq. \eqref{eq:baseResistivity} leads to
simple expressions
at a macroscopic level \citep{tikhonchuk2024}
by relating the current and heat flux,
\begin{eqnarray}
{\bf j}_e & \equiv &  -  e \frac{ 4\pi}{3}   
\int_0^\infty d\varv_e \varv_e^3 {\bf {f}_{e,1} }  \ ,  \\
{\bf q}_{e} & \equiv & \frac{  4\pi  }{3} 
\int_0^{\infty} d\varv_e \, 
\varv_e^3 \varepsilon_e
{\bf f_{1,e}}  \ ,
\end{eqnarray}
with generalized forces, as 
\begin{eqnarray}
\label{eq:fluxJ-force}
{\bf j}_e & = & \sigma_{e \, ||}^{SB} \, {\bf E}_{eff} + \alpha_{e,J \, ||}^{SB}  \, \nabla_{\bf x} (k_BT_e) \ , \\
\label{eq:fluxQ-force}
{\bf q}_e & = & - \alpha_{e,Q \, ||}^{SB}  \, k_BT_e {\bf E}_{eff} - \chi_{e \, ||}^{SB}   \, \nabla_{\bf x} (k_BT_e) \ ,
\end{eqnarray}
where ${\bf E}_{eff} \equiv {\bf E} + \left . \nabla_{\bf x} (n_ek_BT_e) \, \right / (en_e)$ is the effective electric field.
In Eqs. \eqref{eq:fluxJ-force} and \eqref{eq:fluxQ-force},
the electrical and thermal conductivities due to the stimulated Bremsstrahlung effect write
\begin{eqnarray}
 \sigma_{e \, ||}^{SB}  =  \frac{e^2n_e}{m_e  \overline{\nu}_{ei}^{SB}  } \quad , \quad 
  \chi_{e \, ||}^{SB}   =  \frac{9}{2}  \frac{n_ek_BT_e}{m_e  \overline{\nu}_{ei}^{SB} }\ ,
\end{eqnarray}
while the thermoelectric and Peltier coefficients are defined as 
\begin{eqnarray}
\alpha_{e,J \, ||}^{SB}    =  \frac{1}{2} \frac{en_e}{m_e  \overline{\nu}_{ei}^{SB}} \quad , \quad 
  \alpha_{e,Q \, ||}^{SB}   =  3 \frac{en_e}{m_e  \overline{\nu}_{ei}^{SB}} \ .
\end{eqnarray}
Assuming a quasi-neutral plasma yields a zero-current Eq. \eqref{eq:fluxJ-force}, which allows us to rewrite Eq.
\eqref{eq:fluxQ-force} in the standard form,
\begin{eqnarray}
{\bf q}_e  =  - \kappa_{e \, ||}^{SB}   \, \nabla_{\bf x} (k_BT_e) \ ,
\end{eqnarray}
with the thermal conductivity 
\begin{eqnarray}
\kappa_{e \, ||}^{SB} = 3 \frac{n_ek_BT_e}{m_e   \overline{\nu}_{ei}^{SB}} \ .
\end{eqnarray}
The stimulated Bremsstrahlung conductivities can be compared to the classical Spitzer ones
(that includes the effect of both the electron--ion and electron--electron Coulomb collisions) for a fully ionized hydrogen plasma 
\citep{Spitzer1953,Kuritsyn2006,huba2016nrl}:
\begin{eqnarray}
\sigma_{e \, ||}^{SH}  =   \frac{1}{0.51 } \frac{e^2 n_e}{ m_e \overline{\nu}^{C,SH}} \quad , \quad
\kappa_{e \, ||}^{SH}  =  3.2 \frac{n_ek_BT_e}{m_e \overline{\nu}^{C,SH}} \ ,
\end{eqnarray}
where $ \overline{\nu}^{C,SH} = \left ( 2/ \pi \right )^{1/2} \nu^C_{ei}(\varv_{e}^{th}) / 3 $.
The conductivity ratio can then be expressed as a function of a single parameter, $\left . \overline{\nu}^{C,SH} \right /  \overline{\nu}_{ei}^{SB}$, as
\begin{eqnarray}
\label{eq:sigmaSBoverSH}
\left .  \sigma_{e \, ||}^{SB} \right / \sigma_{e \, ||}^{SH} &  =& 0.51  \left . \overline{\nu}^{C,SH} \right /  \overline{\nu}_{ei}^{SB} \ , \\
\left .   \kappa_{e \, ||}^{SB}\right /  \kappa_{e \, ||}^{SH} &  =& (3/3.2) \left . \overline{\nu}^{C,SH} \right /  \overline{\nu}_{ei}^{SB} \ .
\end{eqnarray}
These ratios are plotted in Fig. \ref{fig:conductivity_ratio} in a temperature--density diagram.
The spanned values are relevant
for the variety of plasmas in the solar corona.
Some values of specific interest are highlighted. They correspond to  observations of 
 typical X-ray jets  \citep{shen2021}, quiet-Sun nanoflares and campfires \citep{Podladchikova2025arXiv}, and selected large flares.
The luminous, soft X-ray-emitting plasma in accretion disks of active galactic nuclei \citep{Igarashi2024}, 
typical of Seyfert galaxies, is also indicated. 
These brightening events have a common feature: they reach
the range of temperature and density  corresponding to a low (that is, below unity)
frequency ratio of $\left . \overline{\nu}^{C,SH} \right /  \overline{\nu}_{ei}^{SB}$, 
which indicates that stimulated Bremsstrahlung
scattering dominates Coulomb scattering at thermal electron energies.\\

\begin{figure}[h!]
    \centering
    \includegraphics[width=\hsize]{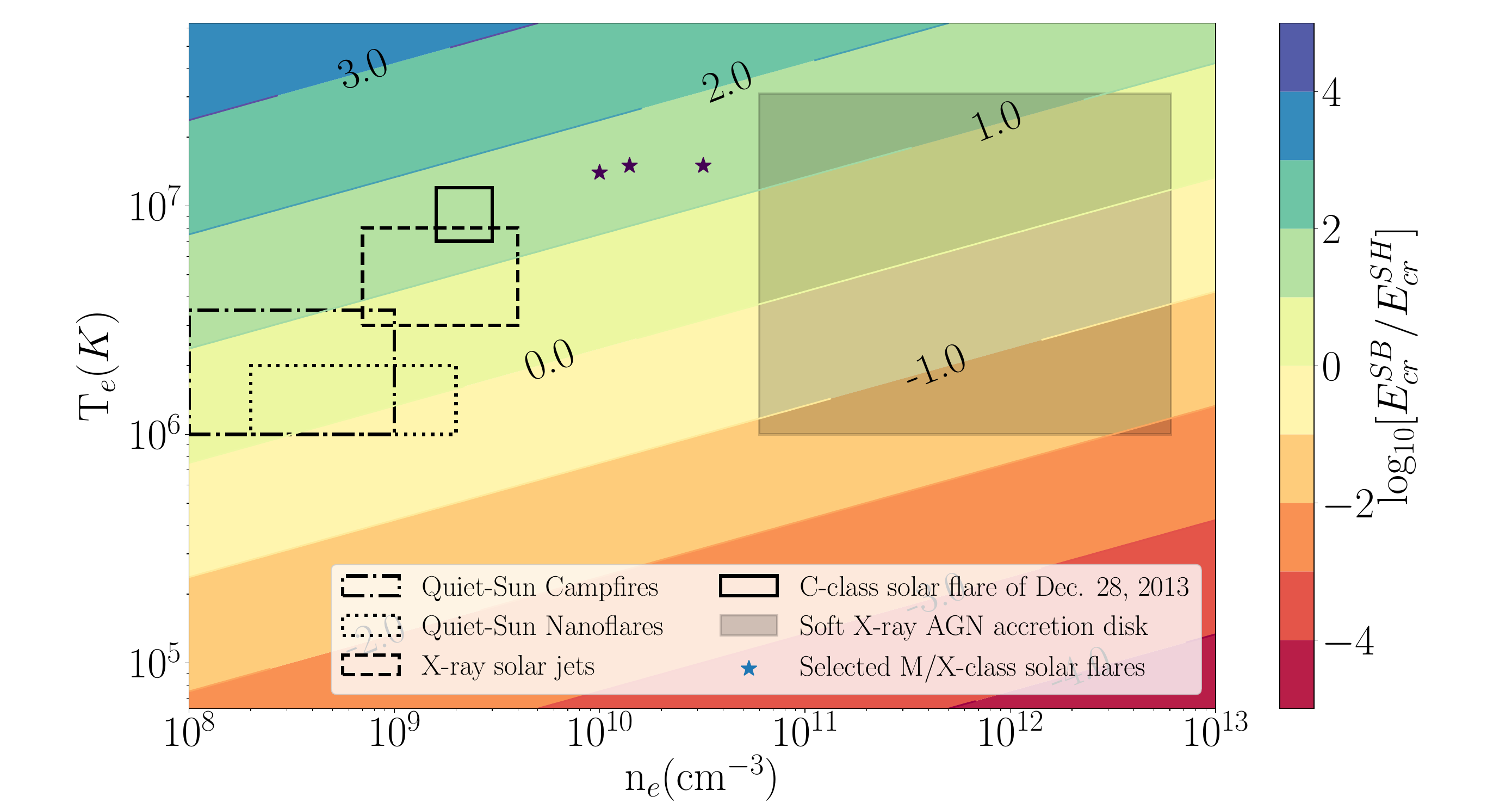}
       \caption{Critical electric field due to stimulated Bremsstrahlung, $E^{SB}_{cr}$, normalized to the Dreicer electric field
       and shown in logarithmic scale as a function of the plasma density and temperature.
       The frequency ratio  $  \overline{\nu}_{ei}^{SB}  / \overline{\nu}^{C,SH} $ was
       evaluated as indicated in Fig. \ref{fig:conductivity_ratio}. 
       Various astrophysical plasmas are represented with symbols (rectangles or stars), which are the same as in Fig. \ref{fig:conductivity_ratio} }
          \label{fig:dreicer}
    \end{figure}

Further, stimulated Bremsstrahlung suppresses runaway electrons due to a weaker dependence on electron velocity.
The Dreicer electric field 
characterizes this process.
It can be computed by considering the momentum equation for electrons
for a space homogeneous plasma and by inserting the drifting Maxwellian velocity distribution
\begin{eqnarray}
M_e^d =  n_e \left ( \frac{m_e}{2\pi k_B T_e} \right )^{3/2}
\exp \left ( - \frac{m_e ({\bf v}_e - {\bf u}_e )^2}{2k_B T_e} \right ) \ ,
\end{eqnarray}
in the electron--ion friction force, yielding  \citep{Dreicer1959}
\begin{eqnarray}
\label{eq:qtemvtDreicer0}
m_e \frac{\partial {\bf u}_e}{\partial t} + e {\bf E}
& = &{\bf F}_{ei} \ ,
\end{eqnarray}
where the friction force writes
\begin{eqnarray}
{\bf F}_{ei}
& = & - \frac{m_e}{n_e} \int_{\mathbb{R}^3} d^3 \varv_e {\bf v}_e \nu_{ei}^{SB} (\varv_e) M_e^d  \\
& = & - e \, E^{SB}_{cr} \, \varv_e^{th}  \frac{ {\bf u}_e }{ u_e^3} \left [ 
 \sqrt{\frac{2}{\pi}}
 u_e 
\exp \left ( - 
u_e^2/ (2 \, {\varv_e^{th}}^2 ) 
\right ) 
 \right . \nonumber \\
& +  & \left .  \left . 
\, \varv_{e}^{th}
\left (
u_e^2/ {\varv_e^{th}}^2 -1
\right ) \,  \mbox{erf} \left (
u_e / ( \sqrt{2} \, \varv_e^{th} )
\right )
 \right ] \right .
 \ . \nonumber
\end{eqnarray}
In the latter expression, we introduced the electric field
\begin{eqnarray}
\label{eq:elec_field_cr}
E^{SB}_{cr} =
\frac{8}{3}   \sqrt{\frac{2}{\pi}}  
\frac{m_e  \overline{\nu}_{ei}^{SB} \varv_e^{th}}{e }\ ,
\end{eqnarray}
which can be compared to the Dreicer electric field due to Coulomb collisions \citep{Dreicer1959,Tsap_2024}:
\begin{eqnarray}
E^{SH}_{cr} = 3 \sqrt{\frac{\pi}{2}} \frac{m_e  \overline{\nu}^{C,SH}  \varv_e^{th}}{e }\ .
\end{eqnarray}
The ratio between these electric fields is shown in Fig. \ref{fig:dreicer}.

\begin{figure}[h!]
    \centering
    \includegraphics[width=\hsize]{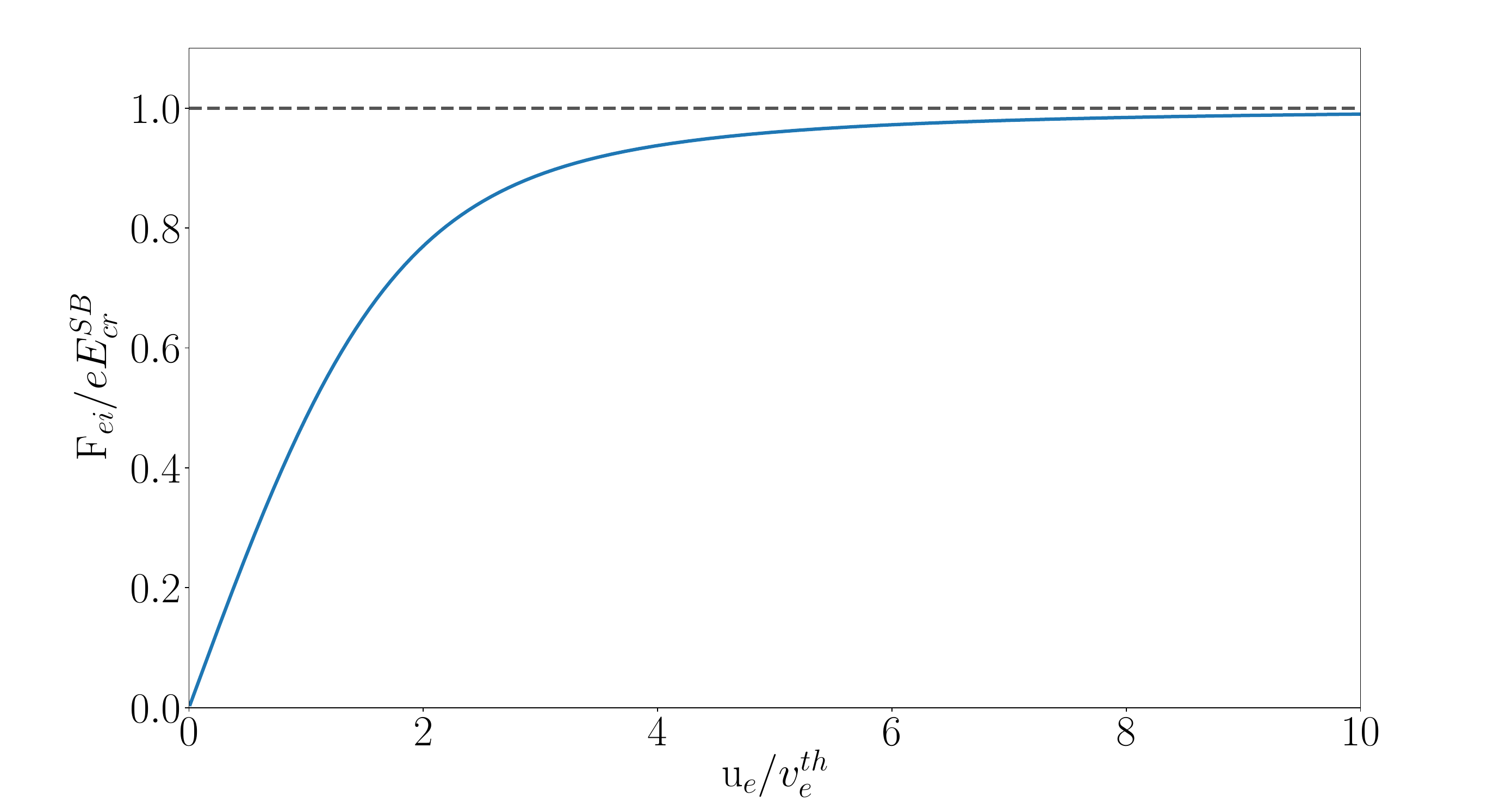}
       \caption{Friction force due to stimulated Bremsstrahlung, normalized with the threshold electric field \eqref{eq:elec_field_cr}, as a function of mean velocity.}
          \label{fig:friction}
    \end{figure}

Assuming a constant and homogeneous electric field, Eq.
\eqref{eq:qtemvtDreicer0} is an autonomous, first-order ordinary differential equation. 
Contrary to the Coulomb case, the friction force on the right-hand side
does not have a bell-shaped dependence
on the mean velocity. Instead, it is monotonous and plateaus at high mean velocities, as shown in Fig. \ref{fig:friction}.
Therefore, if the electric field is
below the threshold $E^{SB}_{cr}$, the friction force dominates and the plasma decelerates.
Moreover, runaway electrons are not produced.
This can be seen by examining the motion of a single electron,
\begin{eqnarray}
m_e \frac{\partial {\bf v}_e}{\partial t} + e {\bf E}
& = & {\bf F}_{ei} ( {\bf v}_e ) \ ,
\end{eqnarray}
where the friction force,
\begin{eqnarray}
\label{eq:frictionforceSB}
{\bf F}_{ei} ( {\bf v}_e ) = -m_e {\bf v}_e \nu_{ei}^{SB}  (\varv_e) = - \frac{ {\bf v}_e }{ \varv_e } eE_{cr}^{SB} \ ,
\end{eqnarray}
does not depend on the electron-velocity modulus. This stands in contrast to the friction force due to  Coulomb scattering,
which vanishes at high velocity, and gives a critical velocity above which electrons become runaways \citep{Gurevich1994}.
If both processes coexist, the critical velocity exists only for above-threshold
electric fields, $E>E_{cr}^{SB}$,
in which case the critical velocity 
is increased by a factor of $ \sqrt{ E / (E - E_{cr}^{SB})  } $
with respect to the pure Coulomb case.
The consequence is that, for
typical large flare plasmas, the accelerating electric field, $E$, should exceed the classical Dreicer electric field, $E_{cr}^{SH}$, by more than one order of magnitude to allow 
electrons to be accelerated (only linearly), as shown in Fig. \ref{fig:dreicer}. In this case, 
due to the shifted critical velocity the number of these
electrons should also be much smaller
than what is classically considered.

\section{Comparison with solar-flare observations}
\label{sec:application}

In the flaring solar corona plasma, 
stimulated Bremsstrahlung scattering can be more efficient than Coulomb collisions
at scattering both fast and thermal electrons, as illustrated by Fig. \ref{fig:conductivity_ratio}.
Such a regime corresponds to a large frequency ratio,
$ \overline{\nu}_{ei}^{SB}  / \overline{\nu}^{C,SH} $, which is maximal for the hottest and less-dense plasmas considered in this figure (upper left corner
of  Fig. \ref{fig:conductivity_ratio}).
This leads us to discuss how this process could influence the different models of return current in coronal loops.
Further, a specific return-current model that relies on stimulated Bremsstrahlung
was built from the transport coefficients of Sect. \ref{sec:transportcoeffs}. It is
used to reinterpret several observations.

\subsection{Return-current models in solar-flare loops}

Once accelerated at the flare loop top by the reconnection process \citep{Li2022}, 
energetic electrons propagate down along the magnetic-field lines.
The charge separation electric field produced by the fast electrons drives the return current of slow electrons. 
A steady state
can be achieved if the drag force is sufficiently strong to equilibrate the drive field
(the return current is then adjusted quasi-instantaneously to neutralize the charge carried by beam current), 
a situation corroborated 
by observations \citep{Battaglia}.\\

Several stationary return-current models
have been proposed for the transport of energetic electrons 
in the magnetic loops of the solar corona \citep{Battaglia,Xu,Alaoui2021}.
Some of them involve a runaway-electron population that contributes to the return current, due to Coulomb collisions.
The runaway population is produced by a sub-Dreicer electric field and 
could be
transported back in the acceleration region of the flare \citep{Alaoui2021}. 
However, stimulated Bremsstrahlung dominates Coulomb collisions in the flaring-Sun plasma and likely suppresses such runaway electrons;
electrons in the tail of the distribution function experience a non-vanishing friction force, while the critical electric field is typically above the Dreicer electric field; see Fig. \ref{fig:dreicer}.
Strong Langmuir turbulence \citep{Rowland1985} could also accelerate runaway electrons.
However, such acceleration may be hindered by the resisitivity 
due to stimulated Bremsstrahlung scattering.
Indeed, the self-consistent electric field of an electron beam can reduce or suppress Langmuir turbulence
\citep{McClements1989,Zharkova2011}.
This electric field was computed,  for the solar flaring plasma,
by considering the classical resistivity \citep{Spitzer1953} instead of the enhanced resistivity
due to stimulated Bremsstrahlung scattering,  $ 1/ \sigma_{e \, ||}^{SB}$, which is shown in the left panel of Fig. \ref{fig:conductivity_ratio}. 
The overall effect of this enhanced resistivity
on runaway acceleration by Langmuir turbulence may be significant, but it remains to be quantified.\\
Similar to runaways, bouncing fast electrons at the loop footpoints due to magnetic mirroring
may also contribute to the return current, albeit without energy loss \citep{Karlicky1993SoPh}.
This back-scattered component in the return current may be enhanced by the large-angle deflections due to stimulated Bremsstrahlung.
\\ 

We consider, in the next section, electron transport in solar flares  based on 
a simple, stationary model of return current that neglects runaway and back-scattered electrons and their influence on the acceleration site. 
In the radiation-dominated regime,
stimulated Bremsstrahlung creates a stronger friction force than Coulomb collisions,
which is sufficient to establish the steady state with a zero-net current.

\subsection{Anomalous resistivities evidenced by hard X-ray spectra in two flare events}
\label{sec:reinterpretation}

There is evidence that large observed currents of fast electrons, transported in solar-flare magnetic loops, 
induce return currents that build up self-consistently to maintain a quasi-neutral steady state \citep{Battaglia}.
Both the fast electron current, $j_{b \, ||}$,  
and the resistive electric field associated with the return current, $E_{||}$, 
can be deduced from the hard X-ray spectra along the loop and at its end points.
However, the classical Spitzer transport theory 
was found to be inconsistent with measures of the
RHESSI satellite \citep{Battaglia}, as it does not allow us to satisfy the quasi-neutral condition 
$j_{b \, ||} = j_{r \, ||}^C $, where
$j_{r \, ||}^C=\sigma_{e \, ||}^{SH} E_{||}$ is the return current due to Coulomb collision; the discrepancy 
is of at least one order of magnitude. 

An enhanced anomalous resistivity due to an ion--acoustic turbulence
has been proposed to reconcile theory and observations \citep{Xu}.
This approach relies on a calibration of the turbulence level
and neglects the Landau damping of ion--acoustic waves. 
The latter is strong in plasmas with equal electron and ion temperatures 
\citep{Melrose1986,McClements1989}. Hence, the approach implicitly amounts to 
considering different electron and ion temperatures \citep{Russell2025}.\\
Here, we focused on anomalous resistivity with an origin different from the ion--acoustic turbulence,
namely stimulated Bremsstrahlung scattering.
To do so, we analyzed and connected three flare events with very similar characteristics; see Table 
\ref{table:flareparameters}. The occurrence of a return current with anomalous resistivity was demonstrated for two of these events, which were recorded on October 24, 2003 and July 13, 2005. 
Meanwhile, an above-loop-top hard X-ray signal was recorded in the flare event on August 24, 2002. The latter was interpreted as resulting from high precipitation rates of fast electrons, which are strongly scattered by
 stimulated Bremsstrahlung on ions \citep{Duclous}.
 
 The plasma parameters of the three flares are very close to each other.
Consequently, these flares should have a similar infrared part of the microwave-radiation spectrum as well; 
see the discussion on microwave-radiation transfer by \cite{Duclous}.
Therefore, we consider that these three events have the same frequency ratio---between stimulated Bremsstrahlung scattering and Coulomb collisions---at thermal energies.
This ratio is found to lie in the  $\overline{\nu}_{ei}^{SB}/ \overline{\nu}^{C,SH}  \in  [70-140]  $ range,
as deduced from Fig. \ref{fig:hprofileCollFreq}.
 Here, the uncertainty comes from the different density and temperature profiles, 
 relating the Sun's surface to the flare's loop top,
 on which the microwave radiation transfer was realized.\\
 Next, the return current due to stimulated Bremsstrahlung scattering, $j_{r \, ||}^{SB}$, can be expressed as 
 a function of
 $j_{r \, ||}^{C}$ as
 \begin{eqnarray}
 \frac{j_{r \, ||}^{SB}}{j_{b \, ||}}= \frac{\sigma_{e \, ||}^{SB} }{\sigma_{e \, ||}^{SH}} 
 \frac{j_{r \, ||}^{C}}{j_{b \, ||}} \ ,
 \end{eqnarray}
where the ratio, $j_{r \, ||}^{C} / j_{b \, ||}$, between return and beam currents was estimated by \cite{Xu}
for the flare events recorded on October 24, 2003 and July 13, 2005, while 
the conductivity ratio, $\sigma_{e \, ||}^{SB} / \sigma_{e \, ||}^{SH}$, is known from Eq. \eqref{eq:sigmaSBoverSH}.
A fully ionized corona with $\ln \Lambda_{ei} \sim 20$ \citep{Somov2012PartI} was assumed to obtain
\begin{eqnarray}
\label{eq:estimateF2003}
 \frac{j_{  r \, || }^{SB}}{j_{  b \, ||  }} \in 
 [ 0.055 - 0.26 ]
 \end{eqnarray}
for the flare event on October 24, 2003 and
\begin{eqnarray}
\label{eq:estimateF2005}
 \frac{j_{  r \, || }^{SB}}{j_{ b \, || }} \in 
 [ 0.013 - 0.075 ]
 \end{eqnarray}
 for the one on July 13, 2005. These straightforward estimates need to be tempered, because 
they do not take into account the neglected back-scattered fast electrons in the return current
that may increase the current ratio, $j_{r \, ||}/j_{b \, ||}$.
Also, a low-energy cutoff of  $\varepsilon_{cut}=20$ keV was used to infer
 the fast electron current, $j_{b \, ||}$, although the latter is very sensitive to this cutoff.
The current scales as $j_{b \, ||} \propto \varepsilon_{cut}^{1-\delta}$, where $\delta$ is the measured spectral index of fast electrons \citep{Battaglia,Xu}.
Detailed studies show that higher cutoff energies are admissible, while lower energies, of
around $10$ keV or less, can be discarded \citep{Sui2007,Battaglia}.
For instance, $\varepsilon_{cut}=30$ keV seems a plausible value.
It is also supported by simulated two-component (fast electrons/return current) 
steady-state electron distributions in the flaring plasma \citep{Zharkova2010}.\\
If the return current is evaluated with both an electric field from observations and an electrical conductivity due to Coulomb collisions, such a higher cutoff energy would be detrimental, as it would pull the 
beam current away from the quasi-neutral prescription, $j_{r \, ||}/j_{b \, ||}=1$, as pointed out by \cite{Battaglia}. 
Thus, the Coulomb scattering does not bring enough collisionality to build up a neutralizing current.
Stimulated Bremsstrahlung brings the missing collisionality. 
This allowed us to recover a zero-net current,
if a higher cutoff energy is considered, or if a substantial contribution to the return current is due to back-scattered electrons, for instance.
Hence, the resistivity due to stimulated Bremsstrahlung scattering
has the potential to give the observed anomalous resistivity in solar-flare loops 
self-consistently (that is, without free parameter).

This possibility may be assessed via a statistical analysis. 
A step in that direction
was made by \cite{Alaoui2017} on the basis of data from  RHESSI.
The dataset consisted of 19 flares with breaks in their X-ray spectra, which were interpreted with return-current models.
The plasma temperature of these flares was measured in the 
$17-46$ MK range, while the background density was measured in the $5 \times 10^9 - 3 \times 10^{11} \, \mbox{cm}^{-3}$ range.
Also, an assumed Ohm's law allowed the authors to obtain
fitted nonclassical resistivity: $\sigma_{e ||} / \sigma_{e ||}^{SH} \in [10^{-4} - 10^{-1}]$.
Although a flare-to-flare analysis would be insightful,
the observational constraints brought by \cite{Alaoui2017} seem consistent with the resistivity given by stimulated Bremsstrahlung, 
as shown in Fig. \ref{fig:conductivity_ratio}. 

\begin{table}
\caption{
\label{table:flareparameters} Plasma parameters of flares under study \citep{Battaglia,Xu,Minoshima2011ApJ73111M,KARLICKY2004383}.}
\begin{center}
\begin{tabular}{|c||c|c|c|}
\hline 
Date & Aug. 24, 2002 & Oct. 24, 2003 & July 13, 2005 \\
\hline
$T_e$ & $14 \; \text{MK}$ & $15 \; \text{MK}$ & $15 \; \text{MK}$ \\
\hline
 $n_e$ & $ 10^{10} \; \text{cm}^{-3}$ & $ 1.4\times 10^{10} \; \text{cm}^{-3}$ & $3.2\times 10^{10} \; \text{cm}^{-3}$ \\
\hline
\end{tabular}
\end{center}
\end{table}

\subsection{Thermal-conduction suppression evidenced from coronal seismology}
\label{subsec:thermalcond}

Plasma cooling by thermal conduction is considered to be important in 
the transition layer between the corona and the chromosphere \citep{West2008,Reville2022}. It can also be important 
in the super-hot corona plasma undergoing magnetic reconnection \citep{Oreshina2011,Sharykin2015}. 
Presumably, high Knudsen numbers, $K_n = \varv_{e}^{th}/\nu_{ei}^C L >1$ ($L$ is a characteristic temperature gradient length),
are common in such situations. Nonlocal electron conduction, mediated by
thermal runaway electrons \citep{Gurevich1979}, may be considered
 \citep{Diakonov1988,West2008,Silva2018,Allred_2022}.
However, increased electron collisionality due to stimulated Bremsstrahlung scattering
may result in a much smaller Knudsen number and local heat transport.
Thermal conduction is also largely diminished in
the flaring solar plasma, compared to the Coulomb case, as shown in Fig. \ref{fig:conductivity_ratio}.
Suppression of thermal conduction was evidenced from coronal seismology by \cite{Wang2015}.
Based on a parametrized linear MHD model, the authors found a suppression factor of at least three with respect to the Spitzer conduction, for a flare loop with slow-mode, in-phase oscillations of electron temperature and density.
Several studies by \cite{Wang2018}, \cite{Wang2019}, and \cite{Prasad2021} consolidated the result 
by using nonlinear MHD models, showing that
thermal conduction should be considered negligible in the hot flaring loop \citep{Ofman2022}.

Consistently, stimulated Bremsstrahlung inhibits thermal conduction in the flaring plasma.
This can be quantified for the slow-mode, longitudinal oscillations of 
XUV intensity disturbances observed by \cite{Wang2015}, which are resolved temporally and spatially along the hot flaring loop.
The deduced electron temperature and density profiles are found to oscillate close to the fundamental mode in the 
$T_e = 7-12$ MK, $n_e= 1.6-3 \times 10^9 \ \mbox{cm}^{-3}$ ranges.
 In these plasma conditions, 
 represented with the bold rectangle in Fig. \ref{fig:conductivity_ratio},
 stimulated Bremsstrahlung suppresses
 thermal conduction
 by a (conservative) factor of $\sim 30$ with respect to the Spitzer conduction.

\section{Conclusion}
\label{sec:conclusion}

Electron transport can be strongly modified under the action of a low-frequency radiation field with a high brightness temperature.
We built a set of transport coefficients to account for this effect and exhibited its peculiarities
in the case of flaring events in the solar corona, considering the Sun microwave radiation. 
In this context, the theory was tested against a number of observations, 
in different channels, which support the model.

The radiation-dominated plasma is characterized by a reverse, nonclassical ordering between the
Bremsstrahlung and Coulomb collision frequencies at thermal energies.
One standing plasma of this kind is the flaring solar corona, in which 
the electron mean-free-path becomes independent of its velocity. 
The electron conductivities are 
strongly reduced, while the runaway effect is suppressed.
It is expected that these effects may be most pronounced in super-hot corona plasmas at reconnection sheets, possibly
enhancing the reconnection rates \citep{Schiavo2024,Talbot2024} and governing the transport of electrons escaping from the acceleration site.

The proposed stimulated process is active at all scales in the flaring plasma. At the kinetic scale, 
 large-angle  Bremsstrahlung scattering of electrons 
 can dominate the precipitation rate. It could be modeled numerically with
 particle-in-cell or Vlasov-Fokker-Planck codes to 
study the reconnecting region. Simulations have already demonstrated an 
important role played by the pitch-angle electron scattering \citep{Egedal2012NatPh}.
Hybrid codes \citep{Arnold2019,Drake2019} used to bridge the gap between scales may include the effect as well.
At a global scale, that is, representative of the flaring solar corona, including the radiation-dominated transport coefficients 
in MHD codes
\footnote{This can be done by approximating 
the collision frequency \eqref{eq:simplenueis}, which couples
the radiation field to the transport coefficients. 
Such an approximation can be made 
analytically in the solar corona by prescribing the spectral slope, $\ell^*$. Alternatively, one may use,
online, an appropriate quadrature formula.
At leading order, the accuracy of the coupling is determined by securing a sufficient resolution in the infrared divergence
(that is, the scaling with the inverse of the photon energy)
of both the transport cross-section and photon occupation number, $n_{\nu}$.
}
will allow them to describe conditions where both radiation and Coulomb collisions contribute to the plasma transport properties.

\begin{acknowledgements}
The authors wish to thank Thomas Farges for helping to link the teams of this collaboration.  
A.S. Brun and A. Strugarek acknowledge CNES Solar Orbiter and CNRS/INSU ATST support.
A. Strugarek acknowledges support by the French Agence Nationale de la Recherche (ANR) project STORMGENESIS \#ANR-22-CE31-0013-01.
The authors are thankful to the Whole Sun program (ERC Syg \#810218) running at Institut Pascal in March 2026,
for hosting useful discussions that have improved the content of the paper.
The text has been revised and improved thanks to the anonymous referee, to whom the authors are grateful.
\end{acknowledgements}

\bibliographystyle{aa}
\bibliography{preprint}

@article{Alaoui2017,

doi = {10.3847/1538-4357/aa98de},

url = {https://doi.org/10.3847/1538-4357/aa98de},

year = {2017},

month = {dec},

publisher = {The American Astronomical Society},

volume = {851},

number = {2},

pages = {78},

author = {Alaoui, Meriem and Holman, Gordon D.},

title = {Understanding Breaks in Flare X-Ray Spectra: Evaluation of a Cospatial Collisional Return-current Model},

journal = {\apj}

}

@article{Alaoui2021,
doi = {10.3847/1538-4357/ac0820},
url = {https://dx.doi.org/10.3847/1538-4357/ac0820},
year = {2021},
month = {aug},
publisher = {The American Astronomical Society},
volume = {917},
number = {2},
pages = {74},
author = {Alaoui, Meriem and Holman, Gordon D. and Allred, Joel C. and Eufrasio, Rafael T.},
title = {Role of Suprathermal Runaway Electrons Returning to the Acceleration Region in Solar Flares},
journal = {\apj}
}

@article{Allred_2022,
doi = {10.3847/1538-4357/ac69e8},
url = {https://dx.doi.org/10.3847/1538-4357/ac69e8},
year = {2022},
month = {may},
publisher = {The American Astronomical Society},
volume = {931},
number = {1},
pages = {60},
author = {Allred, Joel C. and Kerr, Graham S. and Gordon Emslie, A.},
title = {Solar Flare Heating with Turbulent Suppression of Thermal Conduction},
journal = {\apj}
}

@article{Arnold2019,

    author = {Arnold, H. and Drake, J. F. and Swisdak, M. and Dahlin, J.},

    title = {Large-scale parallel electric fields and return currents in a global simulation model},

    journal = {Phys. Plasmas},

    volume = {26},

    number = {10},

    pages = {102903},

    year = {2019},

    month = {10},

    issn = {1070-664X},

    doi = {10.1063/1.5120373},

    url = {https://doi.org/10.1063/1.5120373},

    eprint = {https://pubs.aip.org/aip/pop/article-pdf/doi/10.1063/1.5120373/15612415/102903_1_online.pdf},

}

@article{Avrett_2008,
        doi = {10.1086/523671},
        url = {https://doi.org/10.1086%2F523671},
        year = 2008,
        month = {mar},
        publisher = {{IOP} Publishing},
        volume = {175},
        number = {1},
        pages = {229--276},
        author = {Eugene H. Avrett and Rudolf Loeser},
        title = {Models of the Solar Chromosphere and Transition Region from {SUMER} and {HRTS} Observations: Formation of the Extreme-Ultraviolet Spectrum of Hydrogen, Carbon, and Oxygen},
        journal = {ApJS}

}

@ARTICLE{Bahauddin2021NatAs,
       author = {{Bahauddin}, Shah Mohammad and {Bradshaw}, Stephen J. and {Winebarger}, Amy R.},
        title = "{The origin of reconnection-mediated transient brightenings in the solar transition region}",
      journal = {Nat. Astron.},
         year = 2021,
        month = jan,
       volume = {5},
        pages = {237-245},
}

@article{Ballester2018,
        author = {{Ballester, J. L.} and {Carbonell, M.} and {Soler, R.} and {Terradas, J.}},
        doi = {10.1051/0004-6361/201731567},
        journal = {A\&A},
        pages = {A6},
        title = {The temporal behaviour of MHD waves in a partially ionized prominence-like plasma: Effect of heating and cooling},
        url = {https://doi.org/10.1051/0004-6361/201731567},
        volume = 609,
        year = 2018
}

@article{Battaglia,
	author = {Battaglia, M. and Benz, A. O.},
	title = {Observational evidence for return currents in solar flare loops},
	DOI= "10.1051/0004-6361:200809418",
	url= "https://doi.org/10.1051/0004-6361:200809418",
	journal = {A\&A},
	year = 2008,
	volume = 487,
	number = 1,
	pages = "337-344",
}

@BOOK{bekefi,
       author = {{Bekefi}, G.},
        title = "{Radiation Processes in Plasmas}",
         year = 1966,
      publisher = "J. Wiley \& Sons, Wiley Series in Plasma Physics, New York"
    }

@article{Bian2016,

doi = {10.3847/0004-637X/824/2/78},

url = {https://doi.org/10.3847/0004-637X/824/2/78},

year = {2016},

month = {jun},

publisher = {The American Astronomical Society},

volume = {824},

number = {2},

pages = {78},

author = {Bian, Nicolas H. and Kontar, Eduard P. and Emslie, A. Gordon},

title = {SUPPRESSION OF PARALLEL TRANSPORT IN TURBULENT MAGNETIZED PLASMAS AND ITS IMPACT ON THE NON-THERMAL AND THERMAL ASPECTS OF SOLAR FLARES},

journal = {\apj}

}

@ARTICLE{Braginskii1965,
       author = {{Braginskii}, S.~I.},
        title = "{Transport Processes in a Plasma}",
      journal = {Rev. Plasma Phys.},
         year = 1965,
        month = jan,
       volume = {1},
        pages = {205}
}

@article{BROWNING2024,
title = {From kink instability to magnetic reconnection to oscillations in solar flares},
journal = {Fundam. Plasma Phys.},
volume = {10},
pages = {100049},
year = {2024},
issn = {2772-8285},
doi = {https://doi.org/10.1016/j.fpp.2024.100049},
url = {https://www.sciencedirect.com/science/article/pii/S2772828524000141},
author = {Philippa K. Browning and Mykola Gordovskyy and Luiz A.C.A. Schiavo and James Stewart}
}

@article{Chitta2022,
	author = {Chitta, L. P. and Peter, H. and Parenti, S. and Berghmans, D. and Auchère, F. and Solanki, S. K. and Aznar Cuadrado, R. and Schühle, U. and Teriaca, L. and Mandal, S. and Barczynski, K. and Buchlin, É. and Harra, L. and Kraaikamp, E. and Long, D. M. and Rodriguez, L. and Schwanitz, C. and Smith, P. J. and Verbeeck, C. and Zhukov, A. N. and Liu, W. and Cheung, M. C. M.},
	title = {Solar coronal heating from small-scale magnetic braids⋆},
	DOI= "10.1051/0004-6361/202244170",
	url= "https://doi.org/10.1051/0004-6361/202244170",
	journal = {A\&A},
	year = 2022,
	volume = 667,
	pages = "A166",
}

@BOOK{Cowling1957,
       author = {{Cowling}, T.~G.},
        place={New York},
        title = "{Magnetohydrodynamics}",
        year = 1957,
        publisher={Interscience},
}

@ARTICLE{Diakonov1988,
       author = {{D'Iakonov}, S.~V. and {Somov}, B.~V.},
        title = "{Thermal Electrons Runaway from a Hot Plasma during a Flare in the Reverse-Current Model and Their X-Ray Bremsstrahlung}",
      journal = {\solphys},
         year = 1988,
        month = mar,
       volume = {116},
       number = {1},
        pages = {119-139},
}

@article{Drake2019,

    author = {Drake, J. F. and Arnold, H. and Swisdak, M. and Dahlin, J. T.},

    title = {A computational model for exploring particle acceleration during reconnection in macroscale systems},

    journal = {Phys. Plasmas},

    volume = {26},

    number = {1},

    pages = {012901},

    year = {2019},

    month = {01},

    issn = {1070-664X},

    doi = {10.1063/1.5058140},

    url = {https://doi.org/10.1063/1.5058140},

    eprint = {https://pubs.aip.org/aip/pop/article-pdf/doi/10.1063/1.5058140/16135842/012901_1_online.pdf},

}

@article{Dreicer1959,
  title = {Electron and Ion Runaway in a Fully Ionized Gas. I},
  author = {Dreicer, H.},
  journal = {Phys. Rev.},
  volume = {115},
  issue = {2},
  pages = {238--249},
  numpages = {0},
  year = {1959},
  month = {Jul},
  publisher = {American Physical Society},
  doi = {10.1103/PhysRev.115.238},
  url = {https://link.aps.org/doi/10.1103/PhysRev.115.238}
}

@article{Duclous,
    author = {Duclous, R. and Tikhonchuk, V. and Gremillet, L. and Martinez, B. and Leroy, T. and Masson Laborde, P.-E. and Pain, J.-C. and Decoster, A.},
    title = {Radiation-driven diffusive transport of fast electrons in solar flares},
    journal = {Phys. Plasmas},
    volume = {31},
    number = {2},
    pages = {022904},
    year = {2024},
    month = {02},
    issn = {1070-664X},
    doi = {10.1063/5.0162336},
    url = {https://doi.org/10.1063/5.0162336},
    eprint = {https://pubs.aip.org/aip/pop/article-pdf/doi/10.1063/5.0162336/20008734/022904\_1\_5.0162336.pdf},
}

@ARTICLE{Egedal2012NatPh,
       author = {{Egedal}, J. and {Daughton}, W. and {Le}, A.},
        title = "{Large-scale electron acceleration by parallel electric fields during magnetic reconnection}",
      journal = {Nat. Phys.},
         year = 2012,
        month = apr,
       volume = {8},
       number = {4},
        pages = {321-324}
}

@ARTICLE{Gurevich1979,
       author = {{Gurevich}, A.~V. and {Istomin}, Ya. N.},
        title = "{Thermal runaway and convective heat transport by fast electrons in a plasma}",
      journal = {Sov. JETP},
         year = 1979,
        month = sep,
       volume = {50},
        pages = {470}
}

@article{Gurevich1994,
  title = {Runaway electrons in plasma current sheets},
  author = {Gurevich, A. V. and Sudan, R. N.},
  journal = {Phys. Rev. Lett.},
  volume = {72},
  issue = {5},
  pages = {645--648},
  numpages = {0},
  year = {1994},
  month = {Jan},
  publisher = {American Physical Society},
  doi = {10.1103/PhysRevLett.72.645},
  url = {https://link.aps.org/doi/10.1103/PhysRevLett.72.645}
}

@misc{huba2016nrl,
  title={NRL plasma formulary},
  author={Huba, J. D.},
  year={2016},
  publisher={Naval Research Laboratory, Washington, DC: USA}
}

@article{Igarashi2024,

doi = {10.3847/1538-4357/ad4703},

url = {https://doi.org/10.3847/1538-4357/ad4703},

year = {2024},

month = {jun},

publisher = {The American Astronomical Society},

volume = {968},

number = {2},

pages = {121},

author = {Igarashi, Taichi and Takahashi, Hiroyuki R. and Kawashima, Tomohisa and Ohsuga, Ken and Matsumoto, Yosuke and Matsumoto, Ryoji},

title = {Radiation MHD Simulations of Soft X-Ray Emitting Regions in Changing Look AGN},

journal = {\apj}

}

@BOOK{Jackson1962,
       author = {{Jackson}, J},
        title = "{Classical Electrodynamics}",
         year = 1962,
      publisher = "John Wiley and Sons",
    }

@ARTICLE{Karlicky1993SoPh,
       author = {{Karlicky}, Marian},
        title = "{The acceleration of back-scattered beam electrons in a return-current electric field}",
      journal = {\solphys},
         year = 1993,
        month = may,
       volume = {145},
       number = {1},
        pages = {137-150}
}

@article{KARLICKY2004383,
title = {Loop-top gyro-synchrotron source in post-maximum phase of the August 24, 2002 flare},
journal = {New Astron.},
volume = {9},
number = {5},
pages = {383-389},
year = {2004},
issn = {1384-1076},
doi = {https://doi.org/10.1016/j.newast.2004.01.002},
url = {https://www.sciencedirect.com/science/article/pii/S1384107604000144},
author = {Marian {Karlick{\'y}}}
}

@article{Khomenko2020,

    author = {Khomenko, E. and Collados, M. and Vitas, N. and González-Morales, P. A.},

    title = {Influence of ambipolar and Hall effects on vorticity in three-dimensional simulations of magneto-convection},

    journal = {Philos. Trans. R. Soc. A},

    volume = {379},

    number = {2190},

    pages = {20200176},

    year = {2020},

    month = {12},

    issn = {1364-503X},

    doi = {10.1098/rsta.2020.0176},

    url = {https://doi.org/10.1098/rsta.2020.0176},

    eprint = {https://royalsocietypublishing.org/rsta/article-pdf/doi/10.1098/rsta.2020.0176/248711/rsta.2020.0176.pdf},

}

@article{Klimchuk2015,
author = {Klimchuk, James A. },
title = {Key aspects of coronal heating},
journal = {Philos. Trans. R. Soc. A},
volume = {373},
number = {2042},
pages = {20140256},
year = {2015},
doi = {10.1098/rsta.2014.0256},

URL = {https://royalsocietypublishing.org/doi/abs/10.1098/rsta.2014.0256},
eprint = {https://royalsocietypublishing.org/doi/pdf/10.1098/rsta.2014.0256}
}

@article{KochMotz1959,
title = {BREMSSTRAHLUNG CROSS-SECTION FORMULAS AND RELATED DATA},
author = {Koch, H W and Motz, J W},
abstractNote = {A summary is presented on the bremsstrahlung crosssection formulas which are given in a form convenient for practical calculations. Estimates of their accuracy are given for cases where comparisons can be made with experimental results. Correction factors are indicated In either numerical or analytical form. Data pertaining to electron-electron and to thick target bremsstrahlung are discussed. (C.J.G.)},
journal = {Revs. Modern Phys.},
number = {4},
volume = {31},
year = {1959},
month = {10},
pages = {920},
doi = {10.1103/RevModPhys.31.920}
}

@article{Kou2022,
author = {Kou, Yuankun and Cheng, X. and Wang, Yulei and Yu, Sijie and Chen, Bin and Kontar, Eduard and Ding, Mingde},
year = {2022},
month = {12},
pages = {},
title = {Microwave imaging of quasi-periodic pulsations at flare current sheet},
volume = {13},
journal = {Nat. Commun.}
}

@article{Kuritsyn2006,
    author = {Kuritsyn, A. and Yamada, M. and Gerhardt, S. and Ji, H. and Kulsrud, R. and Ren, Y.},
    title = {Measurements of the parallel and transverse Spitzer resistivities during collisional magnetic reconnectiona)},
    journal = {Phys. Plasmas},
    volume = {13},
    number = {5},
    pages = {055703},
    year = {2006},
    month = {05},
    issn = {1070-664X},
    doi = {10.1063/1.2179416},
    url = {https://doi.org/10.1063/1.2179416},
    eprint = {https://pubs.aip.org/aip/pop/article-pdf/doi/10.1063/1.2179416/15802464/055703\_1\_online.pdf},
}

@article{Landi2008,
        doi = {10.1086/527285},
        url = {https://doi.org/10.1086%2F527285},
        year = 2008,
        month = {mar},
        publisher = {{IOP} Publishing},
        volume = {675},
        number = {2},
        pages = {1629--1636},
        author = {E. Landi and F. Chiuderi Drago},
        title = {The Quiet-Sun Differential Emission Measure from Radio and {UV} Measurements},
        journal = {ApJ}

}

@article{Li2022,

doi = {10.3847/1538-4357/ac6efe},

url = {https://doi.org/10.3847/1538-4357/ac6efe},

year = {2022},

month = {jun},

publisher = {The American Astronomical Society},

volume = {932},

number = {2},

pages = {92},

author = {Li, Xiaocan and Guo, Fan and Chen, Bin and Shen, Chengcai and Glesener, Lindsay},

title = {Modeling Electron Acceleration and Transport in the Early Impulsive Phase of the 2017 September 10th Solar Flare},

journal = {\apj}

}

@ARTICLE{Lu2024,
       author = {{Lu}, Zekun and {Chen}, Feng and {Ding}, M.~D. and {Wang}, Can and {Dai}, Yu and {Cheng}, Xin},
        title = "{A model for heating the super-hot corona in solar active regions}",
      journal = {Nat. Astron.},
         year = 2024,
        month = jun,
       volume = {8},
        pages = {706-715},
}

@article{MacBride2022,
doi = {10.3847/1538-4357/ac94c3},
url = {https://dx.doi.org/10.3847/1538-4357/ac94c3},
year = {2022},
month = {oct},
publisher = {The American Astronomical Society},
volume = {938},
number = {2},
pages = {154},
author = {MacBride, Conor D. and Jess, David B. and Khomenko, Elena and Grant, Samuel D. T.},
title = {Ambipolar Diffusion in the Lower Solar Atmosphere: Magnetohydrodynamic Simulations of a Sunspot},
journal = {\apj}
}

@article{mccormick1956,
       author = {{McCormick}, P.~T. and {Keiffer}, D.~G. and {Parzen}, G.},
        title = "{Energy and Angle Distribution of Electrons in Bremsstrahlung}",
      journal = {Phys. Rev.},
         year = 1956,
        month = jul,
       volume = {103},
       number = {1},
        pages = {29-31},
          doi = {10.1103/PhysRev.103.29},
       adsurl = {https://ui.adsabs.harvard.edu/abs/1956PhRv..103...29M}
}

@ARTICLE{McClements1989,
       author = {{McClements}, K.~G.},
        title = "{Langmuir wave generation by thick target electron beams in solar flares - The effects of density variations and reverse currents}",
      journal = {\aap},
         year = 1989,
        month = jan,
       volume = {208},
       number = {1-2},
        pages = {279-286}
}

@book{Melrose1986,
place={Cambridge},
title={Instabilities in Space and Laboratory Plasmas},
publisher={Cambridge University Press},
author={Melrose, D. B.},
year={1986}
}

@ARTICLE{Minoshima2011ApJ73111M,
       author = {{Minoshima}, Takashi and {Masuda}, Satoshi and {Miyoshi}, Yoshizumi and
         {Kusano}, Kanya},
        title = "{Coronal Electron Distribution in Solar Flares: Drift-kinetic Model}",
      journal = {ApJ},
         year = 2011,
        month = may,
       volume = {732},
       number = {2},
          eid = {111},
        pages = {111},
          doi = {10.1088/0004-637X/732/2/111}
}

@article{Nobrega2020,
	author = {Nóbrega-Siverio, D. and Martínez-Sykora, J. and Moreno-Insertis, F. and Carlsson, M.},
	title = {Ambipolar diffusion in the Bifrost code},
	DOI= "10.1051/0004-6361/202037809",
	url= "https://doi.org/10.1051/0004-6361/202037809",
	journal = {A\&A},
	year = 2020,
	volume = 638,
	pages = "A79",
}

@ARTICLE{Nordlund1982,
       author = {{Nordlund}, A.},
        title = "{Numerical simulations of the solar granulation. I. Basic equations and methods.}",
      journal = {\aap},
         year = 1982,
        month = mar,
       volume = {107},
        pages = {1-10}
}

@article{Ofman2022,

doi = {10.3847/1538-4357/ac4090},

url = {https://doi.org/10.3847/1538-4357/ac4090},

year = {2022},

month = {feb},

publisher = {The American Astronomical Society},

volume = {926},

number = {1},

pages = {64},

author = {Ofman, Leon and Wang, Tongjiang},

title = {Excitation and Damping of Slow Magnetosonic Waves in Flaring Hot Coronal Loops: Effects of Compressive Viscosity},

journal = {\apj}

}

@ARTICLE{Oreshina2011,
       author = {{Oreshina}, A.~V. and {Somov}, B.~V.},
        title = "{On the heat conduction in a high-temperature plasma in solar flares}",
      journal = {Astron. Lett.},
         year = 2011,
        month = oct,
       volume = {37},
       number = {10},
        pages = {726-736},
          doi = {10.1134/S1063773711090064},
       adsurl = {https://ui.adsabs.harvard.edu/abs/2011AstL...37..726O}
}

@article{Pascoe_2019,
doi = {10.3847/1538-4357/ab3e39},
url = {https://doi.org/10.3847/1538-4357/ab3e39},
year = {2019},
month = {oct},
publisher = {The American Astronomical Society},
volume = {884},
number = {1},
pages = {43},
author = {Pascoe, D. J. and Smyrli, A. and Van Doorsselaere, T.},
title = {Coronal Density and Temperature Profiles Calculated by Forward Modeling EUV Emission Observed by SDO/AIA},
journal = {\apj}
}

@ARTICLE{Podladchikova2025arXiv,
       author = {{Podladchikova}, O. and {Warmuth}, A. and {Harra}, L. and {Dolla}, L. and {Verbeeck}, C. and {Mierla}, M. and {Rodriguez}, L. and {Parenti}, S. and {Georgoulis}, M.~K. and {Hofmeister}, S.~J. and {Engler}, N. and {West}, M.~J. and {Veronig}, A.~M. and {Antolin}, P. and {Purkhart}, S. and {Long}, D.~M. and {Buchlin}, {\'E}. and {Haberreiter}, M. and {Zhukov}, A.~N. and {Safari}, H. and {Battaglia}, A.~F. and {Soubri{\'e}}, E. and {B{\"u}chel}, V. and {Gissot}, S. and {De Groof}, A. and {Gyo}, M. and {Halain}, J.~P. and {Inhester}, B. and {Kraaikamp}, E. and {M{\"u}ller}, D. and {Pfiffner}, D. and {Rochus}, P. and {Schuller}, F. and {Smith}, P.~J. and {Schmutz}, W. and {Stegen}, K.},
        title = "{Picoflares in the Quiet Solar Corona: Solar Orbiter Observations Halfway to the Sun}",
      journal = {arXiv e-prints},
         year = 2025,
        month = oct,
          eid = {arXiv:2510.10340},
        pages = {arXiv:2510.10340},
          doi = {10.48550/arXiv.2510.10340},
archivePrefix = {arXiv},
       eprint = {2510.10340},
 primaryClass = {astro-ph.SR},
       adsurl = {https://ui.adsabs.harvard.edu/abs/2025arXiv251010340P}
}

@ARTICLE{Prasad2021,
       author = {{Prasad}, Abhinav and {Srivastava}, A.~K. and {Wang}, Tongjiang},
        title = "{Effect of Thermal Conductivity, Compressive Viscosity and Radiative Cooling on the Phase Shift of Propagating Slow Waves with and Without Heating-Cooling Imbalance}",
      journal = {\solphys},
         year = 2021,
        month = jun,
       volume = {296},
       number = {6},
          eid = {105},
        pages = {105}
}

@ARTICLE{Racah1934,
       author = {{Racah}, Giulio},
        title = "{Sopra L'irradiazione Nell'urto di Particelle Veloci}",
      journal = {Il Nuovo Cimento},
         year = 1934,
        month = jul,
       volume = {11},
       number = {7},
        pages = {461-476},
          doi = {10.1007/BF02959918},
}

@ARTICLE{Raouafi2016,
       author = {Raouafi, N.~E. and Patsourakos, S. and Pariat, E. and Young, P.~R. and Sterling, A.~C. and Savcheva, A. and Shimojo, M. and Moreno-Insertis, F. and DeVore, C.~R. and Archontis, V. and T{\"o}r{\"o}k, T. and Mason, H. and Curdt, W. and Meyer, K. and Dalmasse, K. and Matsui, Y.},
        title = "{Solar Coronal Jets: Observations, Theory, and Modeling}",
      journal = {\ssr},
         year = 2016,
        month = nov,
       volume = {201},
       number = {1-4},
        pages = {1-53}
}

@article{Reville2022,
        author = {Réville, V. and Fargette, N. and Rouillard, A. P. and Lavraud, B. and Velli, M. and Strugarek, A. and Parenti, S. and Brun, A. S. and Shi, C. and Kouloumvakos, A. and Poirier, N. and Pinto, R. F. and Louarn, P. and Fedorov, A. and Owen, C. J. and Génot, V. and Horbury, T. S. and Laker, R. and O’Brien, H. and Angelini, V. and Fauchon-Jones, E. and Kasper, J. C.},
        title = {Flux rope and dynamics of the heliospheric current sheet - Study of the Parker Solar Probe and Solar Orbiter conjunction of June 2020},
        DOI= "10.1051/0004-6361/202142381",
        url= "https://doi.org/10.1051/0004-6361/202142381",
        journal = {A\&A},
        year = 2022,
        volume = 659,
        pages = "A110",
}

@article{Rowland1985,
author = {Rowland, H. and Vlahos, L.},
year = {1985},
month = {02},
pages = {},
title = {Return currents in solar flares - Collisionless effects},
journal = {A\&A}
}

@article{Russell2025,

doi = {10.3847/2041-8213/adf74a},

url = {https://doi.org/10.3847/2041-8213/adf74a},

year = {2025},

month = {sep},

publisher = {The American Astronomical Society},

volume = {990},

number = {2},

pages = {L39},

author = {Russell, Alexander J. B. and Polito, Vanessa and Testa, Paola and De Pontieu, Bart and Belov, Sergey A.},

title = {Solar Flare Ion Temperatures},

journal = {\apjl}

}

@book{Rybicki,
      author        = "Rybicki, George B and Lightman, Alan P",
      title         = "{Radiative Processes in Astrophysics}",
      publisher     = "Wiley",
      address       = "New York, NY",
      year          = "1985",
      url           = "https://cds.cern.ch/record/847173",
      doi           = "10.1002/9783527618170",
}

@article{Schiavo2024,

    author = {Schiavo, Luiz A. C. A. and Stewart, James and Browning, Philippa K.},

    title = {The effects of resistivity on oscillatory reconnection and consequences for solar flare quasi-periodic pulsations},

    journal = {Phys. Plasmas},

    volume = {31},

    number = {10},

    pages = {102903},

    year = {2024},

    month = {10},

    issn = {1070-664X},

    doi = {10.1063/5.0226068},

    url = {https://doi.org/10.1063/5.0226068},

    eprint = {https://pubs.aip.org/aip/pop/article-pdf/doi/10.1063/5.0226068/20223523/102903_1_5.0226068.pdf},

}

@article{Sharykin2015,
author = {Sharykin, I. N. and Struminskii, A. B. and Zimovets, I. V.},
year = {2015},
volume = {41},
  issue = {1},
  pages = {53--66},
title = {Plasma heating to super-hot temperatures (>30 MK) in the August 9, 2011 solar flare},
journal = {Astron. Lett.}
}

@article{shen2021,
author = {Shen, Yuandeng },
title = {Observation and modelling of solar jets},
journal = {Proc. R. Soc. A},
volume = {477},
number = {2246},
pages = {20200217},
year = {2021},
doi = {10.1098/rspa.2020.0217},

URL = {https://royalsocietypublishing.org/doi/abs/10.1098/rspa.2020.0217},
eprint = {https://royalsocietypublishing.org/doi/pdf/10.1098/rspa.2020.0217}
}

@article{Silva2018,
	author = {Silva, S. S. A. and Santos, J. C. and Büchner, J. and Alves, M. V.},
	title = {Nonlocal heat flux effects on temperature evolution of the solar atmosphere},
	DOI= "10.1051/0004-6361/201730580",
	url= "https://doi.org/10.1051/0004-6361/201730580",
	journal = {A\&A},
	year = 2018,
	volume = 615,
	pages = "A32",
}

@article{simoesKerr2017,
        author = {Sim\~oes, P. J. A. and Kerr, G. S. and Fletcher, L. and Hudson, H. S. and Gim\'enez de Castro, C. G. and Penn, M.},
        title = {Formation of the thermal infrared continuum in solar flares},
        DOI= "10.1051/0004-6361/201730856",
        url= "https://doi.org/10.1051/0004-6361/201730856",
        journal = {A\&A},
        year = 2017,
        volume = 605,
        pages = "A125"
}

@book{Somov2012PartI,
place={Berlin},
title={Plasma Astrophysics, Part I: Fundamentals and Practice},
publisher={Springer Science \& Business Media},
author={Somov, B. V.},
year={2012}
}

@book{Somov2013PartII,
place={Berlin},
title={Plasma Astrophysics, Part II: Fundamentals and Practice},
publisher={Springer Science \& Business Media},
author={Somov, B. V.},
year={2013}
}

@article{Talbot2024,

doi = {10.3847/1538-4357/ad2a5d},

url = {https://dx.doi.org/10.3847/1538-4357/ad2a5d},

year = {2024},

month = {apr},

publisher = {The American Astronomical Society},

volume = {965},

number = {2},

pages = {133},

author = {Talbot, Jordan and McLaughlin, James A. and Botha, Gert J. J. and Hancock, Mark},

title = {The Effect of Resistivity on the Periodicity of Oscillatory Reconnection},

journal = {\apj}

}

@article{Spitzer1953,
  title = {Transport Phenomena in a Completely Ionized Gas},
  author = {Spitzer, Lyman and H\"arm, Richard},
  journal = {Phys. Rev.},
  volume = {89},
  issue = {5},
  pages = {977--981},
  numpages = {0},
  year = {1953},
  month = {Mar},
  publisher = {American Physical Society},
  doi = {10.1103/PhysRev.89.977},
  url = {https://link.aps.org/doi/10.1103/PhysRev.89.977}
}

@article{Sui2007,
doi = {10.1086/522198},
url = {https://dx.doi.org/10.1086/522198},
year = {2007},
month = {nov},
publisher = {},
volume = {670},
number = {1},
pages = {862},
author = {Sui, Linhui and Holman, Gordon D. and Dennis, Brian R.},
title = {Nonthermal X-Ray Spectral Flattening toward Low Energies in Early Impulsive Flares},
journal = {\apj},
}

@book{tikhonchuk2024,
  title={Particle Kinetics and Laser-plasma Interactions},
  author={Tikhonchuk, V.},
  year={2024},
  publisher={Cambridge Scholars Publishing}
}

@article{Tsap_2024,
doi = {10.1088/1674-4527/ad1bd5},
url = {https://dx.doi.org/10.1088/1674-4527/ad1bd5},
year = {2024},
month = {jan},
publisher = {National Astromonical Observatories, CAS and IOP Publishing},
volume = {24},
number = {2},
pages = {025015},
author = {Tsap, Yu. T. and Stepanov, A. V. and Kopylova, Yu. G.},
title = {Dreicer Electric Field Definition and Runaway Electrons in Solar Flares},
journal = {Res. Astron. Astrophys.}
}

@article{Vogler2005,
	author = {Vögler, A. and Shelyag, S. and Schüssler, M. and Cattaneo, F. and Emonet, T. and Linde, T.},
	title = {Simulations of magneto-convection in the solar photosphere* - Equations, methods, and results of the MURaM code},
	DOI= "10.1051/0004-6361:20041507",
	url= "https://doi.org/10.1051/0004-6361:20041507",
	journal = {A\&A},
	year = 2005,
	volume = 429,
	number = 1,
	pages = "335-351",
}

@article{Wang2015,

doi = {10.1088/2041-8205/811/1/L13},

url = {https://doi.org/10.1088/2041-8205/811/1/L13},

year = {2015},

month = {sep},

publisher = {The American Astronomical Society},

volume = {811},

number = {1},

pages = {L13},

author = {Wang, Tongjiang and Ofman, Leon and Sun, Xudong and Provornikova, Elena and Davila, Joseph M.},

title = {EVIDENCE OF THERMAL CONDUCTION SUPPRESSION IN A SOLAR FLARING LOOP BY CORONAL SEISMOLOGY OF SLOW-MODE WAVES},

journal = {\apjl}

}

@article{Wang2018,

doi = {10.3847/1538-4357/aac38a},

url = {https://doi.org/10.3847/1538-4357/aac38a},

year = {2018},

month = {jun},

publisher = {The American Astronomical Society},

volume = {860},

number = {2},

pages = {107},

author = {Wang, Tongjiang and Ofman, Leon and Sun, Xudong and Solanki, Sami K. and Davila, Joseph M.},

title = {Effect of Transport Coefficients on Excitation of Flare-induced Standing Slow-mode Waves in Coronal Loops},

journal = {\apj}

}

@article{Wang2019,

doi = {10.3847/1538-4357/ab478f},

url = {https://doi.org/10.3847/1538-4357/ab478f},

year = {2019},

month = {nov},

publisher = {The American Astronomical Society},

volume = {886},

number = {1},

pages = {2},

author = {Wang, Tongjiang and Ofman, Leon},

title = {Determination of Transport Coefficients by Coronal Seismology of Flare-induced Slow-mode Waves: Numerical Parametric Study of a 1D Loop Model},

journal = {\apj}

}

@ARTICLE{West2008,
       author = {{West}, M.~J. and {Bradshaw}, S.~J. and {Cargill}, P.~J.},
        title = "{On the Lifetime of Hot Coronal Plasmas Arising from Nanoflares}",
      journal = {\solphys},
         year = 2008,
        month = oct,
       volume = {252},
       number = {1},
        pages = {89-100},
          doi = {10.1007/s11207-008-9243-3},
       adsurl = {https://ui.adsabs.harvard.edu/abs/2008SoPh..252...89W}
}

@ARTICLE{Xie2024,
  
AUTHOR={Xie, Xiaoyan  and Li, Gang  and Reeves, Katharine K.  and Gou, Tingyu },
         
TITLE={Probing turbulence in solar flares from SDO/AIA emission lines},
        
JOURNAL={Front. Astron. Space Sci.},
        
VOLUME={11},

YEAR={2024},

URL={https://www.frontiersin.org/journals/astronomy-and-space-sciences/articles/10.3389/fspas.2024.1383746},

DOI={10.3389/fspas.2024.1383746},

ISSN={2296-987X}}

@article{Xu,
	author = {Xu, L. and Chen, L. and Wu, D. J.},
	title = {Anomalous resistivity in beam-return currents and hard-X ray spectra of solar flares},
	DOI= "10.1051/0004-6361/201220253",
	url= "https://doi.org/10.1051/0004-6361/201220253",
	journal = {A\&A},
	year = 2013,
	volume = 550,
	pages = "A63",
	month = "",
}

@ARTICLE{Yang2020,
       author = {Yang, Zihao and Bethge, Christian and Tian, Hui and Tomczyk, Steven and Morton, Richard and Del Zanna, Giulio and McIntosh, Scott W. and Karak, Bidya Binay and Gibson, Sarah and Samanta, Tanmoy and He, Jiansen and Chen, Yajie and Wang, Linghua},
        title = "{Global maps of the magnetic field in the solar corona}",
      journal = {Science},
         year = 2020,
        month = aug,
       volume = {369},
       number = {6504},
        pages = {694-697}
}

@article{Zharkova2010,
	author = {Zharkova, V. V. and Kuznetsov, A. A. and Siversky, T. V.},
	title = {Diagnostics of energetic electrons
with anisotropic distributions in solar flares - I. Hard X-rays bremsstrahlung emission},
	DOI= "10.1051/0004-6361/200811486",
	url= "https://doi.org/10.1051/0004-6361/200811486",
	journal = {A\&A},
	year = 2010,
	volume = 512,
	pages = "A8",
	month = "",
}

@article{Zharkova2011,
doi = {10.1088/0004-637X/733/1/33},
url = {https://dx.doi.org/10.1088/0004-637X/733/1/33},
year = {2011},
month = {apr},
publisher = {The American Astronomical Society},
volume = {733},
number = {1},
pages = {33},
author = {Zharkova, Valentina V. and Siversky, Taras V.},
title = {THE EFFECTS OF ELECTRON-BEAM-INDUCED ELECTRIC FIELD ON THE GENERATION OF LANGMUIR TURBULENCE IN FLARING ATMOSPHERES},
journal = {\apj}
}

\begin{appendix}

\section{Stimulated Bremsstrahlung scattering frequencies}
\label{app:sbsfreq}
The stimulated Bremsstrahlung scattering frequency has two contributions, $\nu^{SB}_{ei}=\nu^B_{ei}+\nu^{IB}_{ei}$. 
The first one, $\nu^B_{ei}$, corresponds to stimulated emission of photons, that is, where an electron of incident kinetic energy
$\varepsilon_e$
recoils at energy
$\varepsilon_e^o = \varepsilon_e - \varepsilon_\nu $,
where $\varepsilon_\nu=h\nu$ is the photon energy. The second one, $\nu^{IB}_{ei}$, corresponds to photon absorption, that is, where an electron of incident kinetic energy
$\varepsilon_e$
is accelerated at energy
$\varepsilon_e^o = \varepsilon_e + \varepsilon_\nu $.

Assuming a weakly anisotropic background radiation spectrum,
 the Bremsstrahlung scattering frequencies write in the nonrelativistic limit \citep{Duclous}
 \begin{eqnarray}
 \label{eq:nu_eiB} 
 \nu^B_{ei} (\varv_e) & = & 
  \frac{32}{3}
  \frac{\alpha_f Z^2}{\beta_e}  n_i r_0^2 c   \int_{h\nu_{pe}}^{\varepsilon_e} d\varepsilon_\nu \,
   \left(1 + n_\nu \right) \nonumber \\
  & \times &
  \frac{\sqrt{\varepsilon_e/ \varepsilon_e^o}} {\left( \sqrt{\varepsilon_e} + \sqrt{\varepsilon_e^o} \right)^2}
  G \left ( 
  \frac{\sqrt{\varepsilon_e} -  \sqrt{\varepsilon_e^o} }{
   \sqrt{\varepsilon_e} + \sqrt{\varepsilon_e^o} } \right) \,, 
 \\
\label{eq:nu_eiIB}
 \nu^{IB}_{ei} (\varv_e)  & = &
  \frac{32}{3}
  \frac{\alpha_f Z^2}{\beta_e}  n_i r_0^2 c   \int_{h\nu_{pe}}^{\infty} d\varepsilon_\nu \,
   n_\nu 
\nonumber
   \\
  & \times &
  \frac{\sqrt{\varepsilon_e^o/ \varepsilon_e}} {\left( \sqrt{\varepsilon_e^o} + \sqrt{\varepsilon_e} \right)^2}
  G \left ( 
  \frac{\sqrt{\varepsilon_e^o} -  \sqrt{\varepsilon_e} }{
   \sqrt{\varepsilon_e^o} + \sqrt{\varepsilon_e} } \right) 
 \ ,
 \end{eqnarray}
 where
 \begin{eqnarray} 
 \label{eq:transport_kernel}
 G(x) & = & x \left[ (1-x^2)/(2x^2) + \ln x \right] /(1+x)^2 \ .
 \end{eqnarray}
 Here, the collision frequencies \eqref{eq:nu_eiB} and \eqref{eq:nu_eiIB} were slightly modified from \cite{Duclous},
 to take into account a missing factor 2
\footnote{In this paper, the frequencies $\nu^B_{ei}$ and $\nu^{IB}_{ei}$
have been defined by comparing their contributions to the Coulomb one, 
in the equation for the first-order spherical harmonic of the electron distribution function, where all contributions take the same form.
However,
the frequency $\nu_C$ in Eq. (1) 
of \cite{Duclous}
should be replaced by $\nu_C/2$ in the Coulomb contribution of the scattering model,
therefore the Bremsstrahlung frequencies should be rescaled consistently by a factor 2.
However, the influence of this mistake on the results 
is minimal, and the conclusions of this study remain the same.
}.

\section{Parametrization of the upper solar atmosphere}
\label{app:parametrization}
In this section we consider a simple parametrized model of the electron density and temperature profiles in the solar transition layer and corona.
It is based on a modification, similar to the density model of \cite{Pascoe_2019}, of the C7 solar atmosphere from \cite{Avrett_2008}.
The modification operates at the upper heights, $h \geq h_0$, as
 \begin{eqnarray}
 n_e(h) & = & n_{e,0} \exp  \left (  \displaystyle -   
 \left [ \left ( \frac{h_{0}}{h} \right )^p - 1 \right ]  
 \ln \left ( \frac{n_{e}^{\infty}}{n_{e,0}} \right ) \,
  \right )
 \ , \\
 T_e(h)  & = & 
  T_{e,0} \exp  \left (  \displaystyle -   
 \left [ \left ( \frac{h_{0}}{h} \right )^q - 1 \right ]  
 \ln \left ( \frac{T_{e}^{\infty}}{T_{e,0}} \right ) \,
  \right ) \ .
 \end{eqnarray}
 Here, we choose $h_0=2120$ km, a height at which temperature, $T_{e,0}$, and density, $n_{e,0}$, are set from the C7 model. 
 The asymptotic corona density and temperature are denoted by $n_e^{\infty}$ and $T_e^{\infty}$. 
 For a quiet-Sun corona,  we set $n_e^{\infty}=3.6 \times 10^8 \, \mbox{cm}^{-3}$, $T_e^{\infty}=1 $ MK, 
 while for a flare, we set $n_e^{\infty}=10^{10}  \, \mbox{cm}^{-3}$, $T_e^{\infty}=15 $ MK.
 Various density and temperature gradients are considered, with $p,q\in [3,20]$. They
  are illustrated in Fig. \ref{fig:simple_profiles}.
\begin{figure}[h!]
    \includegraphics[width=\hsize]{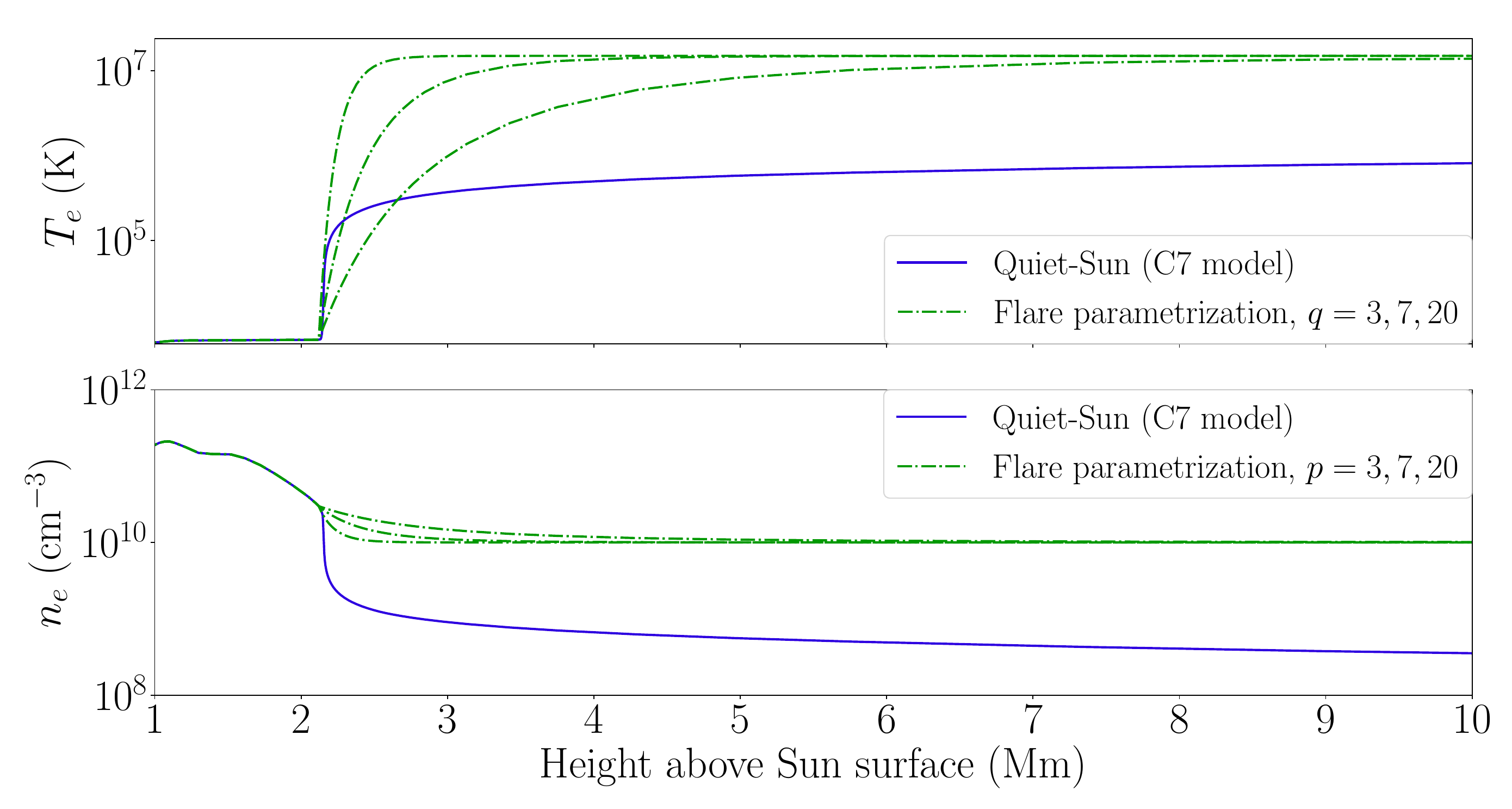}
       \caption{Radial profiles of the electron temperature ({\it top}) and density ({\it bottom})  plotted from a high chromosphere location 
       to a high corona location. The quiet-Sun C7 atmosphere model \citep{Avrett_2008} is represented in solid blue lines.
These profiles are modified with a parametrization from the transition layer to obtain corona  plasma condition approaching that of a flare 
(dashed green lines).
       }
          \label{fig:simple_profiles}
\end{figure}

\section{Magnetized electron transport}
\label{app:magelectras}

In this section we describe how the resistive electric field is modified when the
transport is magnetized, assuming that stimulated Bremsstrahlung scattering dominates over Coulomb scattering or collisions with neutrals.
The obtained structure is analogous to well-established models \citep{Braginskii1965,Ballester2018,Khomenko2020}, although transport coefficients differ.

\subsection{Closure on the current and heat flux}

Equation \eqref{eq:baseResistivity} can be inverted as
\begin{eqnarray}
\label{eq:invertedBaseResistivity}
{\bf f_{e,1} } & = &  \left . {\bf f_{e,1} } \right |_{\chi=0}
+ \frac{\chi}{1+\chi^2} {\bf b} \, \times  \left . {\bf f_{e,1} } \right |_{\chi=0} \nonumber \\
& + & \frac{\chi^2}{1+\chi^2} {\bf b} \, \times  \left [ {\bf b} \, \times    \left . {\bf f_{e,1} } \right |_{\chi=0} \right ] \ ,
\end{eqnarray}
where
$\chi= \nu_{c \, e} / \nu^{SB}_{ei}(\varv_e)$ is the Hall parameter, $\nu_{c \, e} = e|{\bf B}| / (m_ec)$ is the electron cyclotron frequency,
 ${\bf b} = {\bf B} / |{\bf B}|$, and
\begin{eqnarray}
 \left . {\bf f_{e,1} } \right |_{\chi=0} & = & \frac{1}{\nu^{SB}_{ei}(\varv_e)}
 \left [  \frac{ \partial M_e 
 }{\partial \varv_e}  \frac{e{\bf E}}{m_e}  -  \varv_e  \nabla_{\bf x} M_e \right ] \ .
\end{eqnarray}
Considering Eq. \eqref{eq:invertedBaseResistivity} as a basis, 
the coefficients of Eqs. \eqref{eq:fluxJ-force} and \eqref{eq:fluxQ-force}  become tensor-valued due to the magnetic field:
\begin{eqnarray}
\label{eq:fluxJ-forceB}
{\bf j}_e & = & \sigma_{e }^{SB} \, {\bf E}_{eff} + \alpha_{e,J}^{SB}  \, \nabla_{\bf x} (k_BT_e) \ , \\
\label{eq:fluxQ-forceB}
{\bf q}_e & = & - \alpha_{e,Q }^{SB}  \, k_BT_e {\bf E}_{eff} - \chi_{e }^{SB}   \, \nabla_{\bf x} (k_BT_e) \ ,
\end{eqnarray}
where $ \sigma_{e}^{SB}$, $\alpha_{e,J}^{SB}$,  $\alpha_{e,Q}^{SB}$ and $\chi_{e }^{SB}$ are the tensor-valued transport coefficients.
In the following, we denote by $\zeta_e^{SB}$ any of these coefficients.
It can be decomposed, using the standard notations of \cite{Braginskii1965}, as
\begin{eqnarray}
\label{eq:generic_zetaF}
\zeta_{e }^{SB} \, {\bf F} 
& =  &
\zeta_{e \, ||}^{SB} ( {\bf F} \cdot {\bf b} ) \, {\bf b}
+ \zeta_{e \, \perp}^{SB}   {\bf b}  \, \times  \left ( {\bf F}   \, \times  {\bf b} \right )
+ \zeta_{e \, \wedge}^{SB}  {\bf b}  \, \times   {\bf F} \ ,
\end{eqnarray}
where ${\bf F}$ is a generalized force.
The transport coefficient $\zeta_{e }^{SB}$ then takes the form
\begin{eqnarray}
\label{eq:decompositionzeta}
{\zeta}_e^{SB} = \zeta_{e \, ||}^{SB} {\bf I} 
              + ( \zeta_{e \, \perp}^{SB}  - \zeta_{e \, ||}^{SB}) {\bf \Pi} ( {\bf b}) 
              + \zeta_{e \, \wedge}^{SB} {\bf A} ({\bf b}) \ ,
\end{eqnarray}
where ${\bf I} $ is the identity matrix,  ${\bf \Pi} ( {\bf b}) =  {\bf I} - {\bf b} \otimes {\bf b} $ is a projection matrix
and
\begin{eqnarray}
{\bf A} ({\bf b})  = 
\left [ 
\begin{array}[!h]{ccc}
0 & - b_z & b_y \\ 
 b_z & 0 & -b_x \\
 -b_y & b_x  & 0
  \end{array}
\right ]  
\end{eqnarray}
is a skew-symmetric matrix. These matrices satisfy the following relations
${\bf \Pi}^2={\bf \Pi}$, ${\bf A} {\bf \Pi}={\bf \Pi}{\bf A} =
{\bf A}$, ${\bf A}^2 = - {\bf \Pi}$, where the dependence of these matrices on ${\bf b}$ is made implicit, for the sake of lisibility.
The non-parallel components of the tensor-valued transport coefficients depend on the magnetic field. 
They can be expressed with the functions shown in Fig. \ref{fig:nonparalleltransportcoeffs}, that
take values in the interval $[0,1]$ and only depend on the average
Hall parameter,
$\overline{\chi}= \nu_{c \, e} / \overline{\nu}^{SB}_{ei}$,  as
\begin{eqnarray}
\sigma_{e \, \perp}^{SB}   =  \sigma_{e \, ||}^{SB}    \,
\mathcal{S}_{\perp} \left ( \overline{\chi} \right )
\ & ,&  \
 \sigma_{e \, \wedge}^{SB}   =   \sigma_{e \, ||}^{SB}   \,
\mathcal{S}_{\wedge} \left ( \overline{\chi} \right ) \ , \\
\alpha_{e , J \, \perp}^{SB}   =  \alpha_{e , J \, ||}^{SB}  \,
\mathcal{A}_{J \, \perp} \left ( \overline{\chi} \right )
\ & , & \
\alpha_{e , J \, \wedge}^{SB}   =   \alpha_{e , J \, ||}^{SB}   \,
\mathcal{A}_{J \, \wedge} \left ( \overline{\chi} \right )
\ , \\ 
\alpha_{e , Q \, \perp}^{SB}   =   \alpha_{e , Q \, ||}^{SB}  \,
\mathcal{A}_{Q \, \perp}  \left ( \overline{\chi} \right ) 
\ & , & \  
\alpha_{e , Q \, \wedge}^{SB}   =  \alpha_{e , Q \, ||}^{SB}  \,
\mathcal{A}_{Q \, \wedge}  \left ( \overline{\chi} \right )
\ ,    \\ 
\chi_{e \, \perp}^{SB}   =  \chi_{e \, ||}^{SB}   \,
\mathcal{C}_{ \perp}  \left ( \overline{\chi} \right )
\ & , & \
\chi_{e \, \wedge}^{SB}   =  \chi_{e \, ||}^{SB}  \,
 \mathcal{C}_{\wedge}  \left ( \overline{\chi} \right )
\ ,   
\end{eqnarray}
where
\begin{eqnarray}
\mathcal{S}_{\perp} \left ( \overline{\chi} \right ) & = & \frac{1}{2}  \left [ 
 C^3 \, {\overline{\chi} \,  }^{-6} \exp \left (  C \, {\overline{\chi} \,  }^{-2}\right ) E_1  \left (  C \, {\overline{\chi} \,  }^{-2} \right ) \right . \nonumber \\
& + &
 \left . 
C \, {\overline{\chi} \,  }^{-2}  - C^2 \, {\overline{\chi} \,  }^{-4}
\right ] \ ,  \nonumber \\
\mathcal{S}_{\wedge} \left ( \overline{\chi} \right ) & = & \frac{C^3}{2 \overline{\chi}^{6}} 
\left [
- \pi \,  
 \exp \left ( C \, {\overline{\chi} \,  }^{-2} \right ) 
 \mbox{erfc} \left ( C^{1/2} \, {\overline{\chi} \,  }^{-1} \right )
 \right . \nonumber \\
& + & \left . C^{-1/2} \, \overline{\chi} 
\left (
2C^{-5/2} 
\, \overline{\chi}^4 - \pi^{1/2} C^{-1} \overline{\chi}^2 /2 +\pi^{1/2}  \right ) 
\right ] \ , \nonumber \\
\mathcal{A}_{J \, \perp} \left ( \overline{\chi} \right ) & = &
\frac{C}{2 \overline{\chi}^{8}}  \left [ -C^2  \exp \left (  C \, {\overline{\chi} \,  }^{-2}\right ) E_1  \left (  C \, {\overline{\chi} \,  }^{-2} \right ) 
\left ( 5  \overline{\chi}^2
+ 2 C \right  )\right . \nonumber \\ 
& + & \left .  \overline{\chi}^{2}  \left ( 2C^2+3C \overline{\chi}^{2} - \overline{\chi}^{4} \right ) \right ] \ , \nonumber \\
\mathcal{A}_{J \, \wedge} \left ( \overline{\chi} \right ) & = & \frac{C^4}{\overline{\chi}^{8} }
\left [ \pi \,  
 \exp \left ( C \, {\overline{\chi} \,  }^{-2} \right ) 
 \mbox{erfc} \left ( C^{1/2} \, {\overline{\chi} \,  }^{-1} \right )
 \left ( 5  \overline{\chi}^{2}  / (2 C) +1 \right )\right . \nonumber \\
& + & \left . 
4  \overline{\chi}^{5} / (3 C^3) - 16  \overline{\chi}^{3} / (3 C^2) - 8  \overline{\chi} / (3C)  
\right ] \ , \nonumber \\
\mathcal{A}_{Q \, \perp} \left ( \overline{\chi} \right ) & = &  \frac{C^4}{3 \overline{\chi}^{8} } \left [
-  \exp \left (  C \, {\overline{\chi} \,  }^{-2}\right ) E_1  \left (  C \, {\overline{\chi} \,  }^{-2} \right ) / 2
\right . \nonumber \\
& + & \left .   \overline{\chi}^{6} /C^3 -  \overline{\chi}^{4} /(2C^2) +  \overline{\chi}^{2} / (2C)  \right ] \ , \nonumber \\
\mathcal{A}_{Q \, \wedge} \left ( \overline{\chi} \right ) & = & \frac{C^4}{\overline{6 \chi}^{8} }
\left [
\pi \,  
 \exp \left ( C \, {\overline{\chi} \,  }^{-2} \right ) 
 \mbox{erfc} \left ( C^{1/2} \, {\overline{\chi} \,  }^{-1} \right ) 
 \right . \nonumber \\
  & + &  C^{-4} \overline{\chi} \left . \left ( 5 \overline{\chi}^6  - 2 C \overline{\chi}^4  + 4 C^2 \overline{\chi}^2/3 - 8C^3/3 \right )  \right ] \ , \nonumber \\
\mathcal{C}_{ \perp} \left ( \overline{\chi} \right ) & = &   \frac{C}{9 \overline{\chi}^{10} } \left [
 5C^3 \exp \left (  C \, {\overline{\chi} \,  }^{-2}\right ) E_1  \left (  C \, {\overline{\chi} \,  }^{-2} \right ) 
 \left ( \overline{\chi}^{2} + 2C/5  \right )/ 2
\right . \nonumber \\
& + & \left . 
(\overline{\chi}^{2}+C) \overline{\chi}^{2} \left ( \overline{\chi}^{4} -C \overline{\chi}^{2}/2 -C^2 \right )
\right ] \ ,   \nonumber \\
\mathcal{C}_{ \wedge} \left ( \overline{\chi} \right ) & = &
\frac{1}{\overline{9 \chi}^{10} }
\left [
-  \pi C^5 \,  
 \exp \left ( C \, {\overline{\chi} \,  }^{-2} \right ) 
 \right . \nonumber \\ & \times & \left . 
 \mbox{erfc} \left ( C^{1/2} \, {\overline{\chi} \,  }^{-1} \right ) 
 \left ( 5 \overline{\chi}^{2} / (2C) +1  \right )
 \right . \nonumber \\
& + & \left . 
5  \overline{\chi}^9 -   
4 C^2\overline{\chi}^5 /3
+ 16 C^3 \overline{\chi}^3 / 3 
+  8C^4 \overline{\chi} /3 
\right ] \ , \nonumber
\end{eqnarray}
$C=64/(9 \, \pi)$ and $\displaystyle E_1(z) = \int_z^{\infty} dt \exp(-t)/t $ is a special function referred to as 
an exponential integral.
These transport coefficients along the transverse directions, presented in graphical form in figure \ref{fig:nonparalleltransportcoeffs},
are maximized in a weakly magnetized plasma, $\overline{\chi}\lesssim 1$, and strongly decrease in the opposite limit.
The latter corresponds to the strongly magnetized regime, where solar flares operate. The statement is made more precise in what follows.
While the background magnetic field in the solar corona is about $1-4$ Gauss, at a global scale
\citep{Yang2020}, the magnetic field can be enhanced locally and reach hundreds of Gauss in flares.
Let us assume, conservatively, the mere presence of the background magnetic field,
in the flares of Fig \ref{fig:conductivity_ratio}. We obtain $\overline{\chi}>10^3 $. This is  a range where
the dependance of the transport coefficients on $\overline{\chi} \gg 1$ simplifies and can be compared to the classical dependance
\citep{Braginskii1965,Somov2012PartI}, for instance.

    \begin{figure}[h!]
    \resizebox{\hsize}{!}
             {\includegraphics{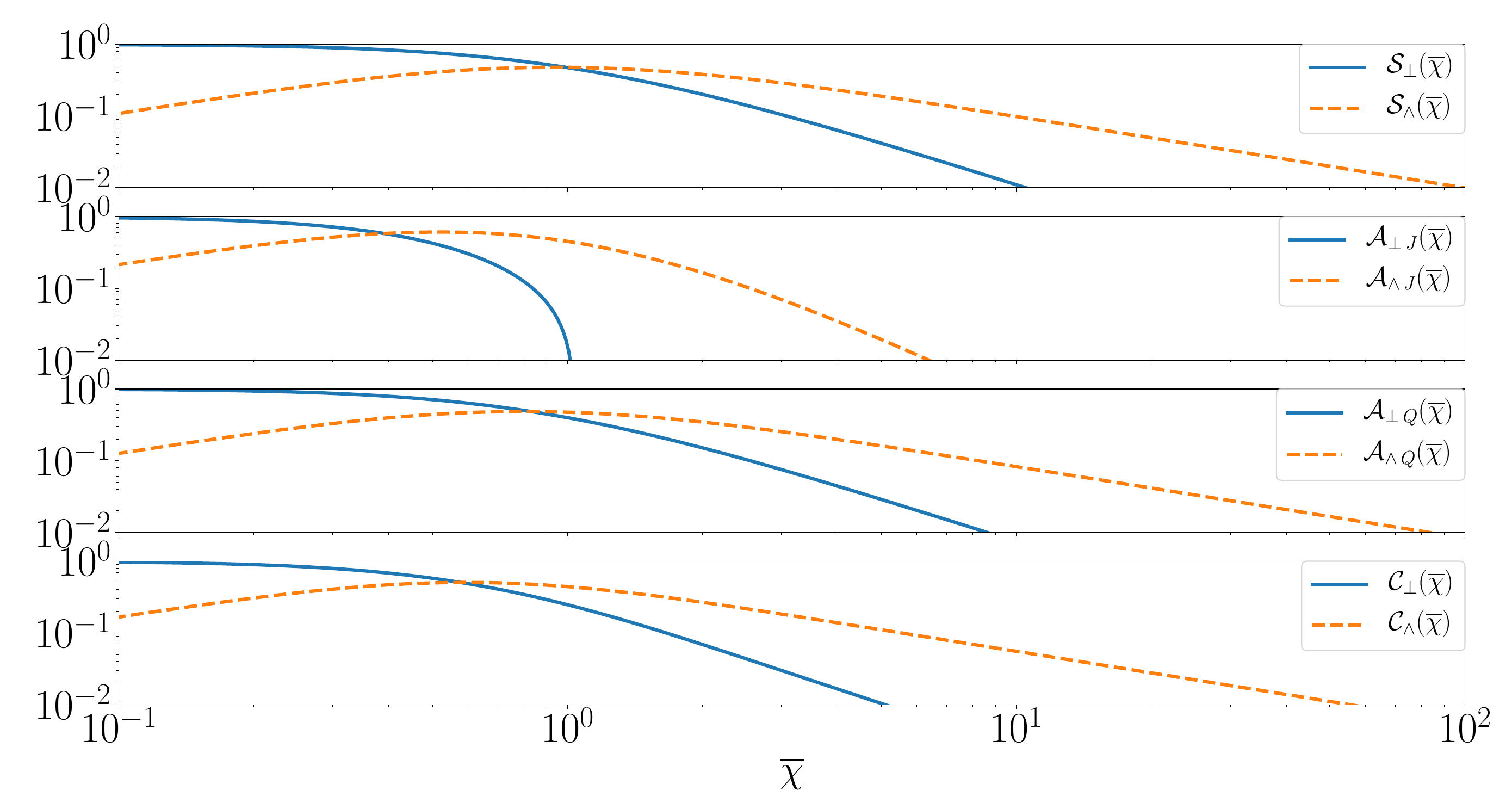}}
       \caption{Nonparallel transport coefficients, normalized to their corresponding parallel transport coefficient.}
       \label{fig:nonparalleltransportcoeffs}
       \end{figure}

\subsection{Resistive electric field}

Let us assume a homogeneous plasma. 
Equation \eqref{eq:fluxJ-forceB} can be recast with resistivity coefficients as
\begin{eqnarray}
\label{eq:eetaj}
{\bf E}
& =  &
\eta_{e \, ||}^{SB} ( {\bf j}_e \cdot {\bf b} ) \, {\bf b}
+ \eta_{e \, \perp}^{SB}   {\bf b} \, \times  \left ( {\bf j}_e  \, \times \, {\bf b} \right )
- \eta_{e \, \wedge}^{SB}  {\bf b} \, \times \,  {\bf j}_e \ ,
\end{eqnarray}
where $\eta_{e \, ||}^{SB} =  1 \left / \sigma_{e \, ||}^{SB} \right . $, 
$\eta_{e \, \perp}^{SB}  =  \sigma_{e \, \perp}^{SB} \left / \, \left [ \left ( \sigma_{e \, \perp}^{SB}  \right )^2 + \left ( \sigma_{e \, \wedge}^{SB}  \right )^2 \right ] \right . $ and\\
$\eta_{e \, \wedge}^{SB}  =  \sigma_{e \, \wedge}^{SB} \left / \, \left [ \left ( \sigma_{e \, \perp}^{SB}  \right )^2 + \left ( \sigma_{e \, \wedge}^{SB}  \right )^2 \right ] \right . $.
The resistive electric field \eqref{eq:eetaj} can be used to describe resistive diffusion,  ambipolar diffusion and Hall transport. 
It takes the compact expression
$ {\bf E} = {\eta}_e^{SB} {\bf j}_e$, by introducing the resistivity tensor
\begin{eqnarray}
{\eta}_e^{SB} = \eta_{e \, ||}^{SB} {\bf I} 
              + ( \eta_{e \, \perp}^{SB}  - \eta_{e \, ||}^{SB}) {\bf \Pi} ( {\bf b}) 
- \eta_{e \, \wedge}^{SB} {\bf A} ({\bf b}) \ ,
\end{eqnarray}
which will be used in the next section.

\subsection{Quasi-neutral thermal conduction}

Assuming a quasi-neutral plasma yields a zero-current Eq.
\eqref{eq:fluxJ-forceB} and allows to rewrite Eq. \eqref{eq:fluxQ-forceB} in the tensorial form of \cite{Braginskii1965}
\begin{eqnarray}
 {\bf q}_e = - \kappa_e^{SB} \nabla_{\bf x} \left  ( k_B T_e \right ) \ ,
 \end{eqnarray}
where the thermal-conductivity tensor writes
\begin{eqnarray}
\kappa_e^{SB} = \chi_e^{SB} - \alpha_{e , Q}^{SB} \eta_e^{SB} \alpha_{e , J}^{SB} \ .
\end{eqnarray}
Here, all transport coefficients can be written 
and composed by using the decomposition
\eqref{eq:decompositionzeta}.

\end{appendix}

\end{document}